\documentclass[aps,prl,groupedaddress,longbibliography]{revtex4-1}
\usepackage{bm,amsfonts,graphicx,cancel}
\usepackage{latexsym,amssymb,amsmath,color}

\usepackage{theorem}
\usepackage{ulem}

\newcommand{\red}[1]{\textcolor{black}{#1}}

\def\<{\langle}
\def\>{\rangle}

\newcommand{\var}[1]{{\sigma^2_{Z}}\left( t \right)}

\begin{document}

\title{Turbulent bifurcations in intermittent shear flows : from puffs to oblique stripes}
\author{Takahiro Ishida$^1$}
\author{Yohann Duguet$^2$}
\author{Takahiro Tsukahara$^1$}

\affiliation{
$^1$Department of Mechanical Engineering, Tokyo University of Science, 278-8510 Chiba, Japan \\
$^2$LIMSI-CNRS, Campus Universitaire d'Orsay, Universit\'e Paris-Saclay, 91405 Orsay, France
}

\date{\today}

\begin{abstract}
Localised turbulent structures such as puffs or oblique stripes are building blocks of the intermittency regimes in subcritical wall-bounded shear flows. These turbulent structures are investigated in incompressible pressure-driven annular pipe flow using direct numerical simulations in long domains. For low enough radius ratio $\eta$, these coherent structures have a dynamics comparable to that of puffs in cylindrical pipe flow. For $\eta$ larger than 0.5, they take the shape of helical stripes inclined with respect to the axial direction. The transition from puffs to stripes is analysed statistically by focusing on the axisymmetry properties of the associated large-scale flows. It is shown that the transition is gradual : as the azimuthal confinement relaxes, allowing for an azimuthal large-scale component, oblique stripes emerge as predicted in the planar limit. The generality of this transition mechanism is discussed in the context of subcritical shear flows. 

\end{abstract}

\maketitle

\section{1. Introduction}

Subcritical wall-bounded shear flows have the ability to sustain turbulent motion despite the linear stability of the laminar regime. As the flow rate is decreased starting from the fully turbulent regime, partial relaminarisation is frequently observed prior to global relaminarisation, leading to the intermittent occurrence of turbulence in an otherwise laminar flow \cite{Manneville2016}. For cylindrical pipe flow driven by either a fixed pressure gradient or a fixed mass flux, this intermittency manifests itself as disordered sequences of so-called \emph{puffs}, i.e. turbulent structures filling the cross-section of the pipe but localised in the streamwise direction \cite{Wygnanski73}. The self-organisation of these trains of puffs near criticality results from the interplay between local relaminarisation events and spatial proliferation \cite{Avila2011,Barkley2016}. For planar shear flows such as plane Couette or plane Poiseuille flow, or any combination of both \cite{Klotz2016}, intermittency manifests usually itself as spatially periodic patterns of laminar-turbulent coexistence in the form of \emph{stripes} of turbulence \cite{Prigent02,Tsukahara2005}. These stripes display a non-zero angle with respect to the streamwise direction. Closer to the onset of turbulence, these oblique structures have been reported to break up into disjoint finite-sized turbulent spots whose interaction raises interesting questions from a phase transition point of view \cite{Duguet2010, Lemoult2016}.  Despite recent progress coming mainly from low-order models \cite{Manneville2016,Barkley11}, a general explanation for these different types of self-organisation is still lacking. \\

Identifying the conditions that lead either to puff-like or stripe-like  structures in a given flow geometry \red{would help to unravel} the mechanisms responsible for the localisation of turbulence. This ambitious task \red{suggests that a flow case should be selected} with a free parameter able to bridge as continuously as possible the two limiting cases of turbulent puffs, on one hand, and turbulent stripes on the other hand. In the context of pressure-driven flows, annular Poiseuille flow (aPf) is an interesting candidate for such a homotopy procedure. This geometry features two long (ideally infinite) co-axial pipes of different radii $R_i$ and $R_o$ $(> R_i)$, between which an incompressible flow is maintained using a fixed axial pressure gradient. This flow geometry is relevant in many important industrial processes ranging from nuclear plants to heat exchangers. As in all pressure-driven flows, the dramatic drop in flow rate associated with the laminar-turbulent transition makes the issues of whether and how turbulence maintains itself important for practical situations. The shape of the associated laminar flow profile depends only on the radius ratio $\eta=R_i/R_o$. \red{This profile turns} out to be linearly stable even for values of the Reynolds number far above those where turbulence starts to be sustained, making the transition in this flow case clearly subcritical for all values of $\eta$ \cite{Walker57,Heaton08}. In a former paper \cite{IDT2016}, the zoology of the different manifestations of laminar-turbulent coexistence was addressed using numerical simulation in moderately long (axially periodic) domains : while helical stripe patterns dominate the low-$Re$ range for $\eta \ge 0.5$, only \red{statistically-axisymmetric} turbulent structures were identified for $\eta=0.1$ and low enough Reynolds number. The former type corresponds to the stripes found in plane Poiseuille flow (pPf) in the presence of a wall curvature that vanishes \red{asymptotically as} $\eta \rightarrow 1^-$. The latter type can be assimilated to puffs as in circular pipe flow, except that the presence of the inner rod implies a different velocity profile both in the laminar and turbulent zones. Simulations for intermediate $\eta=0.3$ featured new axially localised helical structures baptised `helical puffs'. \\

In the present numerical study, we consider all these intermittent regimes in numerical domains longer than in \red{previous studies}, parametrised as before by the radius ratio $\eta$ with emphasis on the patterning property of localized turbulence. Another important control parameter is the Reynolds number ${\rm Re}_\tau = u_\tau d/(2\nu)$, where $u_\tau=\sqrt{\tau/\rho}$ is the friction velocity, with $\tau$ the mean total wall friction proportional to the pressure gradient and $\rho$ is the density of fluid, $\nu$ its kinematic viscosity, and $d=R_o-R_i>0$ measures the spacing between the two cylinders. The details of the numerics are given in Section 2. The dynamics of the flow at marginally low values of ${\rm Re}_\tau$ is analysed in Section 3 for various values of $\eta$ from 0.1 to 0.5. In Section 4, a statistical analysis of the transition from puffs to stripes as $\eta$ increases is suggested, based on the statistical quantification of the large-scale flows present at the laminar-turbulent interfaces. Eventually, the generality of the transition mechanism advanced in Section 4 is discussed in Section 5.

\section{2. Numerical simulation}

\begin{figure}[t]
\begin{center}
\includegraphics[height=25mm]{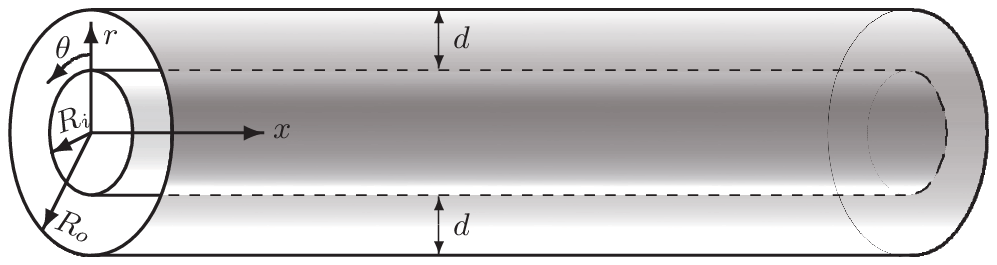} ~~
\includegraphics[height=25mm]{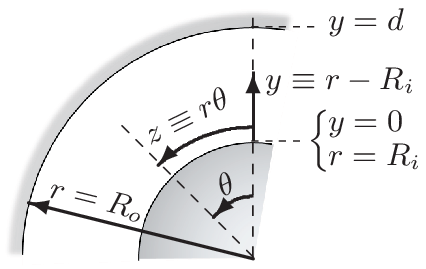}
\end{center}
\caption{Configuration of annular Poiseuille flow and coordinate system.}
\label{fig:std01}
\end{figure}

The geometry is best described using the classical cylindrical coordinate system ($r$, $\theta$, $x$), where the $x$-axis is the axis common to both cylinders. We also define the rescaled azimuthal coordinate $z=r \theta$ and the wall-normal distance from the inner wall $y = r -R_i$. \red{At $r=R_o$ we have $y=d$ and an azimuthal extent $L_{zo}=2\pi R_o$, while for $r=R_i$ we have $y=0$ and $L_{zi}=2\pi R_i$. The specific values of $L_{zo}$ and $L_{zi}$ are listed in Table \ref{tab:para}.} Periodicity is considered in both $x$ and $\theta$, with respective spatial periods $L_x$ and 2$\pi$, while no slip (${\bf u}=0$) is imposed at each wall. The flow between the two cylinders is governed by the incompressible Navier-Stokes equations
\begin{eqnarray}
\nabla \cdot {\bf u} &=& 0, 
\label{cont} \\
{\partial_t {\bf u}}
+ \left( {\bf u} \cdot \nabla \right) {\bf u}
&=& -\rho^{-1} \nabla p + \nu  \nabla^2 {\bf u}.
\label{NS}
\end{eqnarray}
Denoting respectively by $\tau_i$ and $\tau_o$ the wall shear rate at the inner and outer wall, the axial pressure gradient at equilibrium reads ${\rm d} \overline{p} / {\rm d} x = -(2/d)(\tau_o + \eta \tau_i)/(1+ \eta)$, from which $u_\tau=\sqrt{-\rho^{-1}{\rm d} \overline{p} / {\rm d} x}$ and hence ${\rm Re}_\tau$ are defined. The numerical algorithm used to solve Eqs.~(\ref{cont}) and (\ref{NS}) combines fourth-order finite differences  in both $x$ and $\theta$, together with a second-order scheme in $r$ on a non--uniform radial grid. Timestepping is performed using an Adams-Bashforth scheme for the nonlinear terms and a Crank-Nicolson scheme for the wall-normal viscous term, which results in a second-order algorithm. Further information about the numerical method employed here can be found in Refs. \cite{Abe01,IDT2016}. 

All numerical parameters are reported in Table \ref{tab:para}. The local numerical resolution has been checked in \red{our previous article \cite{IDT2016}}. The length of the computational domain has been varied between $51d$ and \red{$180d$} depending on the outcome of the simulations. It is longer than most simulations from Ref.~\cite{IDT2016} in order to capture the interaction between several distinct localised turbulent structures.  \\

\begin{table*}[t]
\caption{Computational conditions for DNS: 
$L_x$ and $L_z$ are the streamwise and azimuthal lengths ($L_{zi} = 2\pi R_i$ and $L_{zo} = 2\pi R_o$); 
$N_x$ and $N_z$ the corresponding grid numbers ($z=r\theta$), while $N_y$ is fixed at 128 for all the cases ($y = r - R_i$). 
Basically, longer $L_x$ with larger $N_x$ were used for DNS at lower ${\rm Re}_\tau$. 
}
\label{tab:para}
\begin{center}
\footnotesize
\begin{tabular}{ccccccc} \hline\hline
$\eta$ & 0.1 & 0.2 & 0.3 & 0.4 & 0.5 & 0.8 \\
${\rm Re}_\tau$                & 46--150    & 50--150   & 50--150   & 52--80   & 52--150  & 52--150  \\\hline
$L_x/d$                      &  51.2--180.0  &  102.4--166.0  &  51.2--160.0  &  102.4--160.0  &  51.2--80.0  &  51.2--80.0  \\
$L_{z i}(\eta) / d$&  0.70 &  1.57  &  2.69  &  4.19  &  6.28  &  25.1   \\
$L_{z o}(\eta) / d$&  6.98  &  7.85  &  8,98  &  10.5  &  12.6  &  31.4   \\
$N_x $ &  2048 or 4096  &  4096  &  2048 or 4096  &  4096  & 2048 &  2048  \\
$N_z $ &  256  &   256  &   512  &   512  & 512 & 1024  \\\hline\hline
\end{tabular}
\end{center}
\end{table*}

\section{3. Temporal dynamics of localised turbulent structures}

The procedure (quenching from the turbulent regime) chosen here is similar to that in Refs.~\cite{PhilipManneville2010,IDT2016} : for each value of $\eta$, a statistically steady turbulent flow is first reached easily by adding a random perturbation of finite amplitude to the laminar base flow at ${\rm Re}_{\tau}=150$. For this value of ${\rm Re}_{\tau}$, turbulence unambiguously occupies the whole numerical domain. Then, ${\rm Re}_{\tau}$ is decreased in finite steps until statistically steady laminar-turbulent coexistence is detected from visualisations, usually around ${\rm Re}_{\tau} \approx 80$. The control parameter is then lowered further in smaller steps until turbulence globally collapses, typically for ${\rm Re}_{\tau} \lesssim 50$. Let us denote by ${\rm Re}^c_\tau = Re^c_\tau(\eta)$ \red{the critical value of} ${\rm Re}_{\tau}$ below which no turbulence is sustained in the long time limit. \red{Throughout this paper, the superscript $(\cdot)^c$ will be used to represent a critical point for sustainment of localized turbulence. However, the values suggested in this paper are not based on a full but costly statistical analysis---like e.g. in Ref.~\cite{Avila2011}---but are estimated from our limited set of simulations.} We describe below the different types of spatio-temporal dynamics encountered around ${\rm Re}^c_{\tau}$ in order of increasing $\eta$. Note again that all simulations have been performed here in longer domains than in Ref.~\cite{IDT2016} and that the spatiotemporal regimes resulting from the interaction between different localised patches of turbulence are more complex than in previous works. All the spatiotemporal diagrams shown in this section involve the streamwise velocity averaged azimuthally $\langle u_x/u_\tau \rangle_\theta = \int_{0}^{2\pi}u_x {\rm d} \theta /2\pi$, evaluated at mid-gap.\\

\subsection{3.1 Localised puffs for $\eta=0.1$}

We describe first the regimes found for low $\eta=0.1$. Laminar-turbulent coexistence is easily visualised using spatiotemporal diagrams of the quantity $\langle u_x/u_\tau \rangle_\theta$ at mid-gap as a function of the streamwise coordinate $x$ and of the time $t$ in units of $d/u_{\tau}$. For easier visualisation, the reduced streamwise coordinate $x'$ is used instead of $x$, where $x'=(x-ct)/d$ with $c$ an adjustable streamwise speed. \red{This speed of the reference frame was adjusted simply by eye in order to cancel out the apparent drift of puffs. It should be noted that the present $c$ does not rigorously correspond to the propagation speed of puffs.}
Figures~\ref{fig:std01}(a-e) show such diagrams for ${\rm Re}_{\tau}=56$, 52, 50, 48 and 46, respectively, with the colourmap adjusted so that the red colour is an indicator of local turbulent motion. For ${\rm Re}_{\tau}=56$, the flow consists of a train of four to five several \red{puffs interacting} via sequences of replication and merging events.  At ${\rm Re}_{\tau}=52$ and 50, the number of puffs interacting drops to two or three, with relaminarisation or replication (``splitting'') events occuring on longer timescales and no merging event. For ${\rm Re}_{\tau}=48$, a single isolated puff survives over the whole observation time, though at several occasions its internal fluctuations bring it close to splitting or to collapsing. For ${\rm Re}_{\tau}=46$, the isolated puff \red{attempts to split} into two parts that both collapse after less than $5d/u_\tau$. This decription is completely consistent with that of turbulent puffs in circular pipe flow in Refs.~\cite{Moxey10,Avila2011,Shimizu2014}. This suggests, though it remains to be properly shown, that the transition in an infinitely long pipe would be continuous, with the critical point Re$^c_{\tau}$ for this value of $\eta$ lying around $48 \pm 1$. \\

\begin{figure}[t]
\begin{center}
\includegraphics[height=80mm]{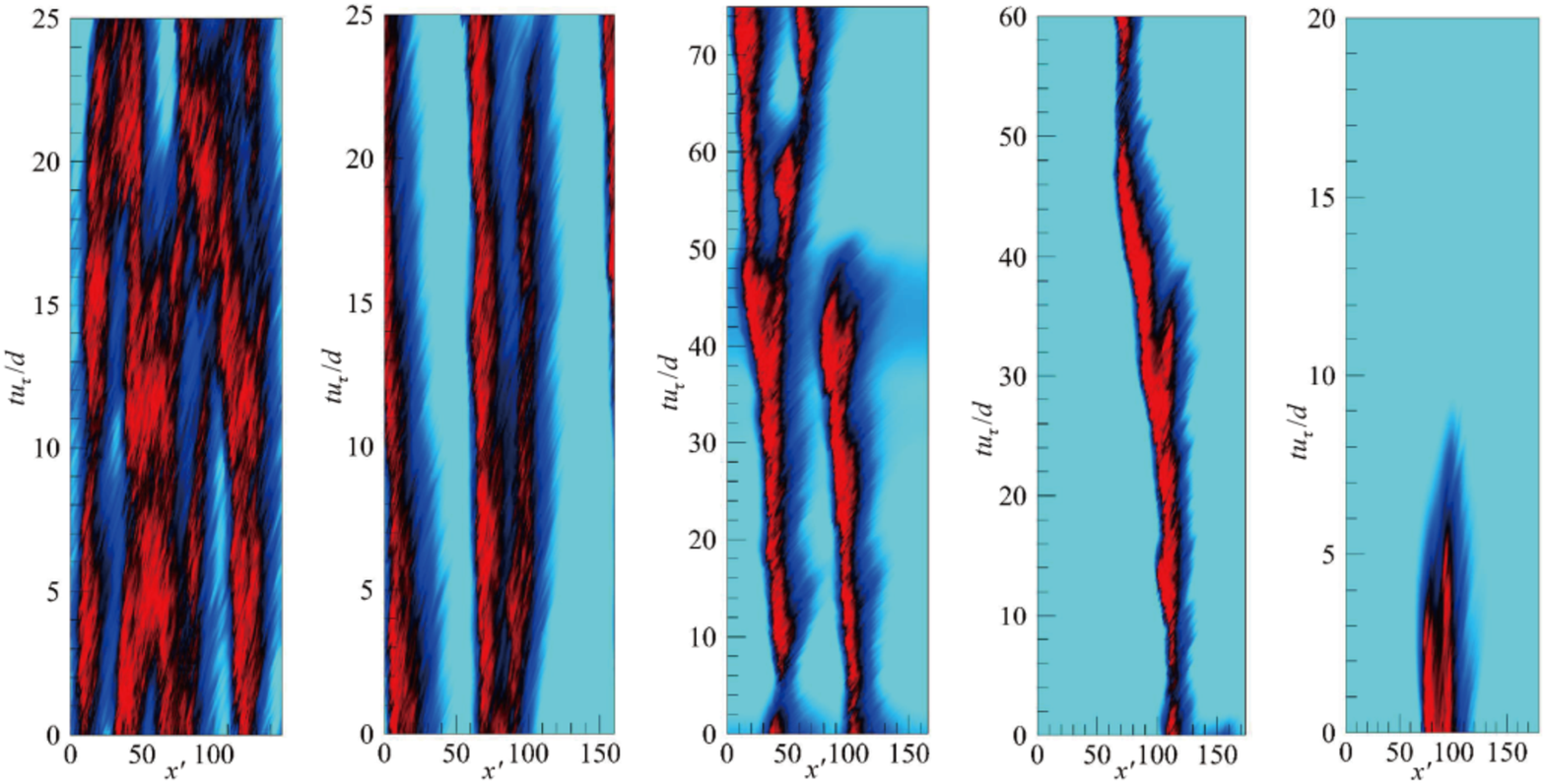}\\
\end{center}
\hspace{0em} (a) Re$_\tau=56$
\hspace{3.3em} (b) Re$_\tau=52$
\hspace{3.9em} (c) Re$_\tau=50$
\hspace{3.5em} (d) Re$_\tau=48$
\hspace{3.5em} (e) Re$_\tau=46$\\
\vspace{0.7em}
\includegraphics[width=8mm, angle =90]{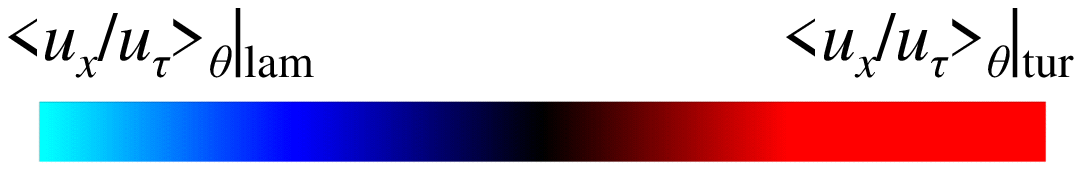}
\vspace{-1.0em}
\caption{Space-time diagram of azimuthally-averaged streamwise velocity at mid-gap for $\eta = 0.1$ and different Re$_\tau$. Colormap of $\langle u_x/u_\tau \rangle_\theta$ is plotted against a reduced variable $x' =(x-ct)/d$, in a moving frame of reference, and its range is ($\langle u_x/u_\tau \rangle_\theta|_{\rm tur}$, $\langle u_x/u_\tau \rangle_\theta|_{\rm lam}$) : (a) (18, 22), (b) (19, 24), (c) (20, 25), (d) (21, 25), and (e) (21, 25).}
\label{fig:std01}
\end{figure}

\begin{figure}[t]
\begin{center}
\includegraphics[width=120mm]{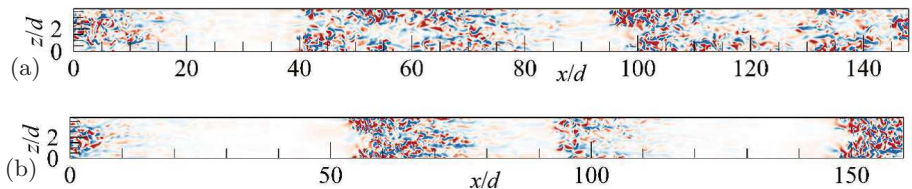}\\
\vspace{-6em} \hspace{-36em}(a)\\
\vspace{ 3em} \hspace{-36em}(b)
\vspace{-0.5em}
\end{center}
\caption{Two-dimensional contours of instantaneous $u'_r$ (blue, red) = ($-0.75$, 0.75) at mid-gap, for $\eta = 0.1$ at Re$_\tau$ = (a) 56 and (b) 52. The mean-flow direction is from left to right.}
\label{fig:2d01}
\end{figure}

Detailed visualisation of the flow structures is achieved using the spatial fluctuations of the radial velocity $u_r'$ around its spatial average in a cylinder at arbitrary $y$. The fluctuating component is defined as :
\begin{eqnarray}
u_i' &=& u_i - \widetilde{u_i}(r, t), \\
\widetilde{u_i}(r, t) &=& \frac{1}{L_x \cdot 2\pi}\iint u_i(x, r, \theta, t){\rm d}x{\rm d}\theta. 
\label{eq:average}
\end{eqnarray}
\red{Note that $\widetilde{u_i}$ does not necessarily equal the laminar flow, because of the presence of fully/localized turbulent region in our  case---in particular, an oblique turbulent region is also accompanied by a large-scale flow also in the azimuthal direction, which may provide non-zero $\widetilde{u_\theta}$ \cite{IDT2016}.}
The flow structures for ${\rm Re}_{\tau}=56$ and 52 are visualised in Fig.~\ref{fig:2d01}. These figures confirm the ressemblance with puffs from circular pipe flow (see e.g. Ref.~\cite{Shimizu2014} for similar visualisations of puffs close to the onset of puff splitting). In particular, each patch of turbulent fluctuations extends over the full cross-section, i.e., \red{they are rather homogeneous} in the $z$ direction with straight laminar-turbulent interfaces. 

\subsection{3.2 Occurrence of helical puffs for $\eta=0.2$ and $0.3$}

\begin{figure}[t]
\begin{center}
\includegraphics[height=80mm]{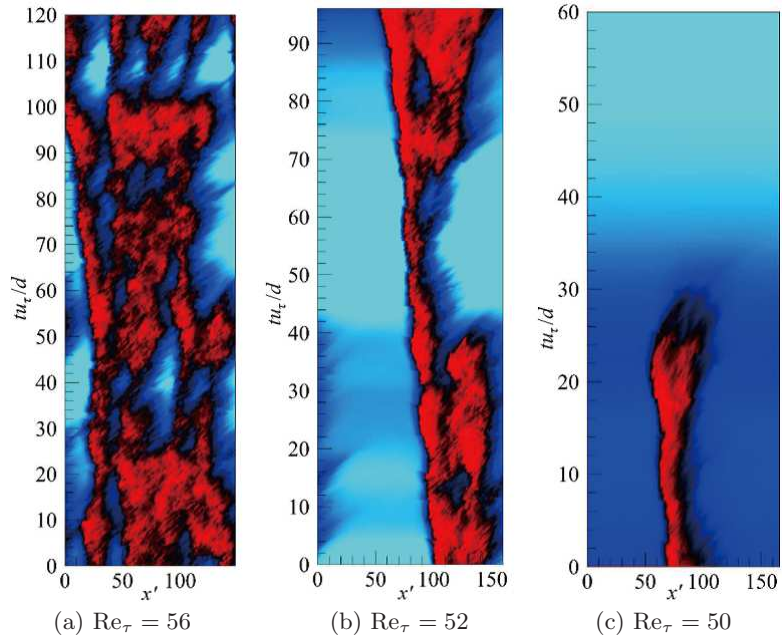}\\
(a) Re$_\tau=56$ \hspace{5em}
(b) Re$_\tau=52$ \hspace{4.5em}
(c) Re$_\tau=50$\\
\vspace{-1.5em}
\end{center}
\caption{Same as Fig.~\ref{fig:std01}, but for $\eta = 0.2$ : ($\langle u_x/u_\tau \rangle_\theta |_{\rm tur}$, $\langle u_x/u_\tau \rangle_\theta |_{\rm lam}$) = (a) (18, 23), (b) (20, 25), and  (c) (20, 27).}
\label{fig:std02}
\end{figure}

\begin{figure}[t]
\begin{center}
\includegraphics[width=120mm]{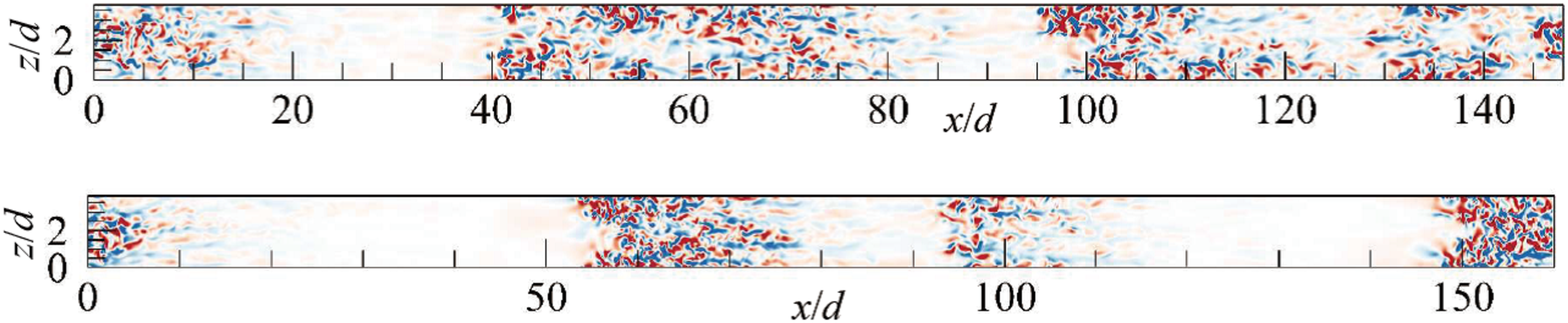}\\
\vspace{-6em} \hspace{-36em}(a)\\
\vspace{ 3em} \hspace{-36em}(b)
\vspace{-0.5em}
\end{center}
\caption{Same as Fig.~\ref{fig:2d01}, but for $\eta = 0.2$ at Re$_\tau$ = (a) 56 and (b) 52.}
\label{fig:2d02}
\end{figure}

Visualisations of the intermittent flow structures and their temporal dynamics for $\eta=0.2$ are shown in Figs.~\ref{fig:std02} and \ref{fig:2d02} using the same quantities as in Figs.~\ref{fig:std01} and \ref{fig:2d01}. The spatiotemporal diagrams for ${\rm Re}_{\tau}=56$, 52, and 50 in Fig.~\ref{fig:std02}(a-c) do not differ much from those for $\eta=0.1$, except perhaps at ${\rm Re}_{\tau}=56$ where the turbulent flow exhibits a stronger tendency toward patterning (emergence of a well-defined streamwise wavelength) than for lower $\eta$. For $\eta=0.2$, the puff sustaining at ${\rm Re}_{\tau}=52$ may split and avoid collapse, while the puff at ${\rm Re}_{\tau}=50$ decays completely, see Figs.~\ref{fig:std02}(b) and (c). This suggests a critical point around ${\rm Re}^c_{\tau}=51\pm1$. The spatial fluctuations of $u'_r(x,z)$ at mid-gap in Fig. \ref{fig:2d02} show a surprising property : some of the laminar-turbulent interfaces identified display obliqueness with respect to the streamwise direction while other do not. 

\begin{figure}[t]
\begin{center}
\includegraphics[height=80mm]{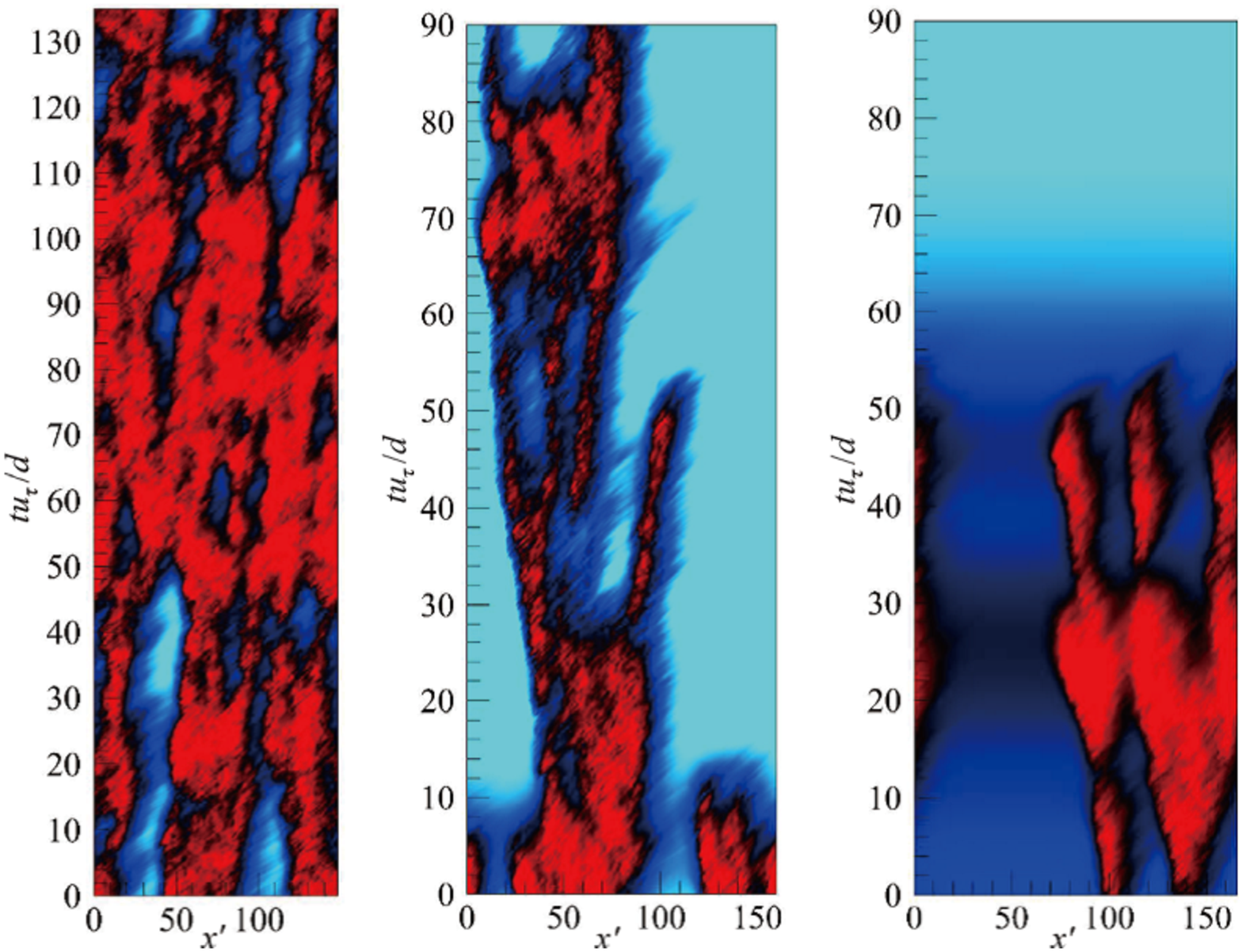}\\
(a) Re$_\tau=56$ \hspace{5em}
(b) Re$_\tau=52$ \hspace{4.5em}
(c) Re$_\tau=50$\\
\vspace{-1.5em}
\end{center}
\caption{Same as Fig.~\ref{fig:std01}, but for $\eta = 0.3$ : ($\langle u_x/u_\tau \rangle_\theta |_{\rm tur}$, $\langle u_x/u_\tau \rangle_\theta |_{\rm lam}$) = (a) (18, 22), (b) (19, 23), and (c) (19, 26).}
\label{fig:std03}
\end{figure}

\begin{figure}[t]
\begin{center}
\includegraphics[width=120mm]{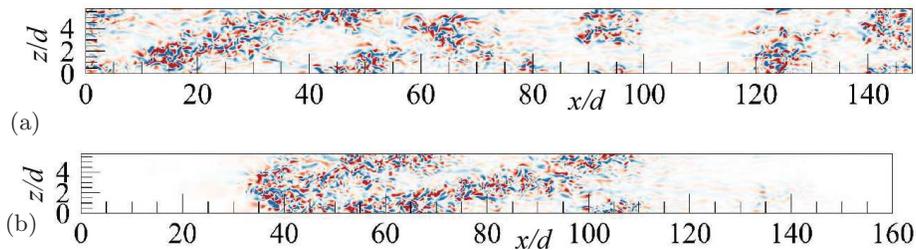}\\
\vspace{-6em} \hspace{-36em}(a)\\
\vspace{ 3em} \hspace{-36em}(b)
\vspace{-0.5em}
\end{center}
\caption{Same as Fig.~\ref{fig:2d01}, but for $\eta = 0.3$ at Re$_\tau$ = (a) 56 and (b) 52.}
\label{fig:2d03}
\end{figure}

Similar data for $\eta=0.3$ confirms the above-mentioned trends, with an even stronger patterning property and a comparable critical value : Fig.~\ref{fig:std03}(c) provides \red{a specific evidence} that, for ${\rm Re}_{\tau}=50$, three puffs separated by a comparable wavelength collapse in synchrony. 
The occurrence of oblique laminar-turbulent interfaces appears also more pronounced, judging from the fluctuations of $u'_r$ shown in Fig.~\ref{fig:2d03}. The resulting intermittent regime for $\eta=0.2$ and 0.3 appears hence as a mixture of both classical straight puffs analogous to those found for $\eta=0.1$, and helical puffs as identified in Ref.~\cite{IDT2016} in shorter domains. \red{We emphasize} the coexistence in space as in time and for a given set of parameters, of both types of structures. This immediately suggests that the transition from (straight) puffs to (oblique/helical) stripes cannot be treated as deterministic, but rather requires a statistical treatment. The statistical analysis to be presented in Section~4.3 will be based on the probablity density function of large-scale azimuthal velocities that is related to the obliqueness of \red{the pattern}.

\subsection{3.3 Occurrence of stripe patterns for $\eta \ge 0.4$}

\begin{figure}[t]
\begin{center}
\includegraphics[width=120mm]{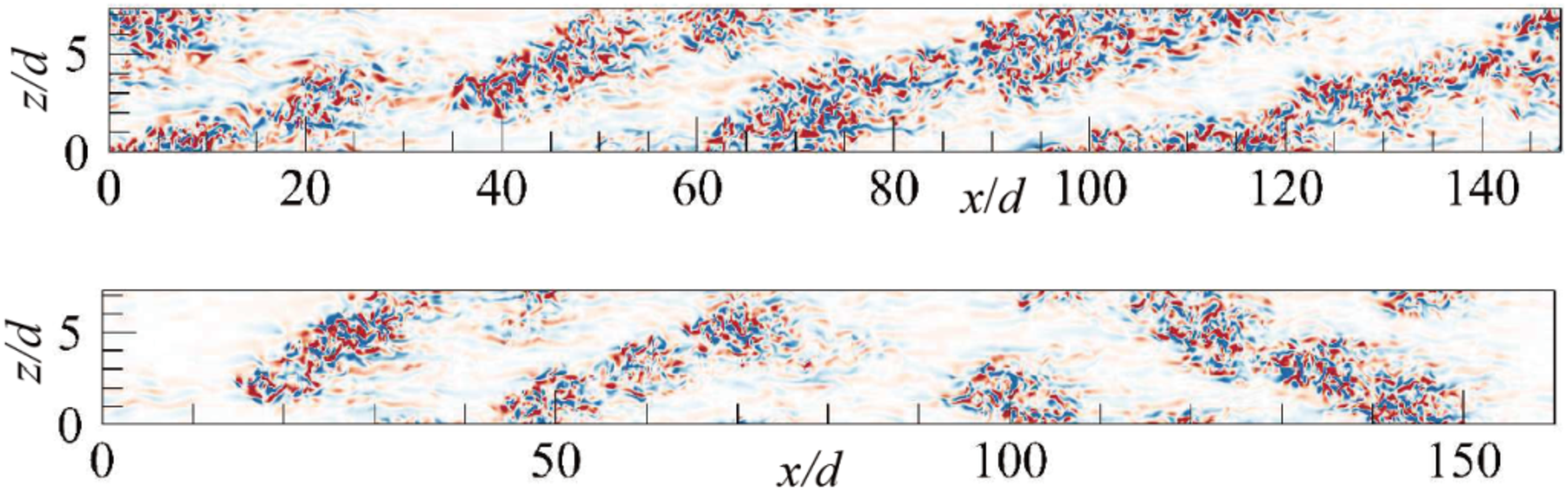}\\
\vspace{-6em} \hspace{-36em}(a)\\
\vspace{ 3em} \hspace{-36em}(b)
\vspace{-0.5em}
\end{center}
\caption{Same as Fig.~\ref{fig:2d01}, but for $\eta = 0.4$ at Re$_\tau$ = (a) 56 and (b) 52.}
\label{fig:2d04}
\end{figure}

\begin{figure}[t]
\begin{center}
\includegraphics[width=120mm]{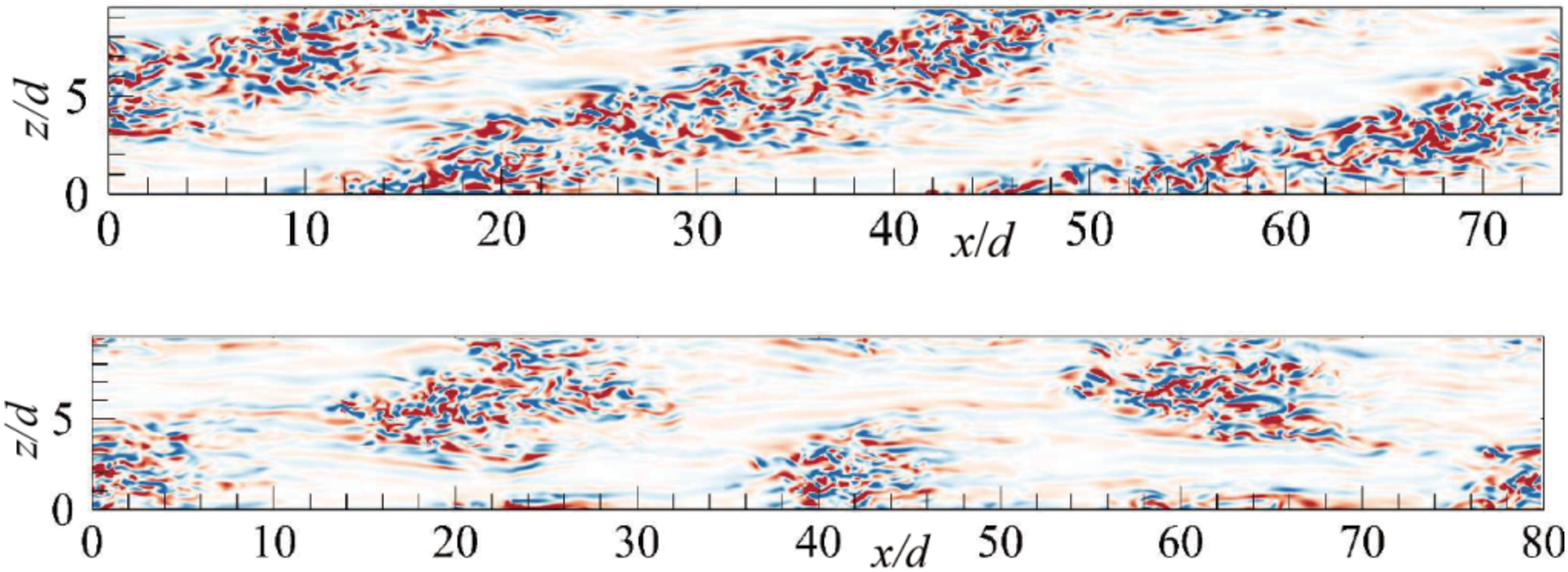}\\
\vspace{-8em} \hspace{-36em}(a)\\
\vspace{ 5em} \hspace{-36em}(b)
\vspace{-0.5em}
\end{center}
\caption{Same as Fig.~\ref{fig:2d01}, but for $\eta = 0.5$ at Re$_\tau$ = (a) 56 and (b) 52.}
\label{fig:2d05}
\end{figure}

Finally, the same indicators have been computed for $\eta=0.4$ and 0.5 at ${\rm Re}_{\tau}=56$ and 52, both apparently above the corresponding value of ${\rm Re}^c_{\tau}$.
The associated spatiotemporal diagrams demonstrate sustained patterning and are not shown. The spatial fluctuations shown for randomly chosen snapshots in Figs.~\ref{fig:2d04} and ~\ref{fig:2d05} show exclusively oblique laminar-turbulent interfaces for ${\rm Re}_{\tau}=56$ and apparently no puff-like structures, suggesting that this corresponds to the periodic stripe regime. \red{In contrast}, ${\rm Re}_{\tau}=52$ shows a sequence of isolated spots \red{less like a} periodic pattern. A similar dynamics was reported in planar cases ($\eta \rightarrow 1$) where for marginally low values of Re$_\tau$ the stripe patterns break up into smaller structures \cite{Xiong15}, and \red{this is a feature specific to} the turbulent regimes slightly above ${\rm Re}^c_{\tau}$.

\section{4. Bifurcation analysis based on large-scale flows}
\label{sec:analysis}

Bifurcations from one turbulent regime to another one are difficult to investigate because, unlike bifurcations of exact steady/periodic states, the presence of turbulent fluctuations both in time and space makes the choice of a well-defined bifurcation parameter non obvious. We intend here to characterise the bifurcation of the \emph{shape} of coherent structures, where no particularly obvious Eulerian indicator emerges to describe for instance the obliqueness of the interfaces. We suggest to link the present study to the limiting planar case $\eta \rightarrow 1$, which has been already discussed in Ref.~\cite{DS2013}, and to generalise it to curved geometries corresponding to $0<\eta<1$. This represents an opportunity to test the limitations of the planar theory in the presence of finite curvature. \red{In addition,} we keep in mind the observation that both types of structures, straight and helical, have been detected for the same parameters at intermediate values of $\eta$. The relevant bifurcation parameter chosen should hence be quantified in a probabilistic manner, with statistics carried out, for each set of parameters $\eta$ and ${\rm Re}_{\tau}$, over both time and space.

\subsection{4.1 Role of the large-scale flows}

For planar flows, it was suggested in Ref.~\cite{DS2013} that the obliqueness of interfaces could be explained qualitatively by the existence of \emph{large-scale flows}. 
In the presence of a sufficiently marked scale separation between small scales (the turbulent fluctuations) and large scales, it was shown analytically that large scales advect the small scales of weakest amplitude. Interfaces between laminar and turbulent motion correspond precisely to the zones where the fluctuations decay from their turbulent amplitude towards zero. It is thus expected that the planar orientation of the interface corresponds precisely to the orientation of the large-scale flow advecting the small scales  at the edges of the turbulent patches. In particular the global angle of the periodic stripe patterns corresponds accurately to the angle of the large-scale flow (cf Fig. 2c in Ref.~\cite{DS2013}) : a non-zero angle is linked with the existence of a \emph{spanwise} component for the large scales. The small scales of interest consist essentially of streaks and streamwise vortices of finite length, which form the minimal ingredients of the self-sustaining process in all wall-bounded shear flows \cite{HKW1995}. The origin of the transverse large scale flow is kinematic rather than dynamic : it can be derived from the mass conservation at the interfaces once streamwise localisation is assumed. In particular the role of the spanwise large-scale component in the planar case is to compensate for the loss of streamwise flow rate inside the turbulent patch (with respect to the reference flow rate inside the laminar zones).

We begin by defining the relevant quantities with \red{notation} adapted to the present cylindrical geometries for any value of $\eta$. \red{Independently of} temporal considerations, large-scale flows ${\bf U}(x,r,\theta)$ can be obtained from any flow field ${\bf u}(x,r,\theta)$ by the application of a low-pass filter $\mathcal{L}$ that selects only the lowest-order modes in $\theta$ and $x$ directions. The spectral cut-off criterion in these two directions is an intrinsic parameter of the filter. The exact choice of the kernel for the low-pass filter (here Heaviside functions in both wave numbers $k_x$ and $k_{\theta}$) matters little as long as the scale separation is sufficiently well pronounced. 
Consider the continuity equation in cylindrical coordinates : 
\begin{eqnarray}
\partial_{x}u_{x} + \frac{1}{r}\partial_{r}(ru_{r}) +  \frac{1}{r}\partial_{\theta}u_{\theta} = 0.
\label{contcyl}
\end{eqnarray}
The divergence operator commutes with $\mathcal{L}$, which leads to the same equation for the large-scale flow : 
\begin{eqnarray}
\partial_{x}U_{x} + \frac{1}{r}\partial_{r}(rU_{r}) +  \frac{1}{r}\partial_{\theta}U_{\theta} = 0 .
\label{contcyl_LSF}
\end{eqnarray}
Multiplying Eq.~(\ref{contcyl_LSF}) by $r$ and integrating it from $r=R_i$ to $R_o$ leads to
\begin{eqnarray}
\int_{R_i}^{R_o}\partial_{\theta}U_{\theta}\,{\rm d}r=- \int_{R_i}^{R_o}\partial_{x}U_{x}r\,{\rm d}r.
\label{eq1}
\end{eqnarray}
We now define radial integration 
 \begin{eqnarray}
\overline{(\cdot)}&=&\frac{\int_{R_i}^{R_o} r (\cdot) \,{\rm d}r}{\int_{R_i}^{R_o} r \,{\rm d}r} .
 \label{def_int}
\end{eqnarray}
Since $\overline{U_x}$ does not vanish, then we can define an angle $\alpha \ge 0$ with respect to the streamwise direction by
\begin{eqnarray}
\tan{\alpha (x, \theta, t)}=\left| \frac{\overline{U_\theta (x, \theta, t)}}{\overline{U_x (x, \theta, t)}} \right|.
\label{def_alpha}
\end{eqnarray}

\subsection{4.2 Spectral analysis}

\begin{figure}[t]
\begin{center}
(a) \includegraphics[width=70mm]{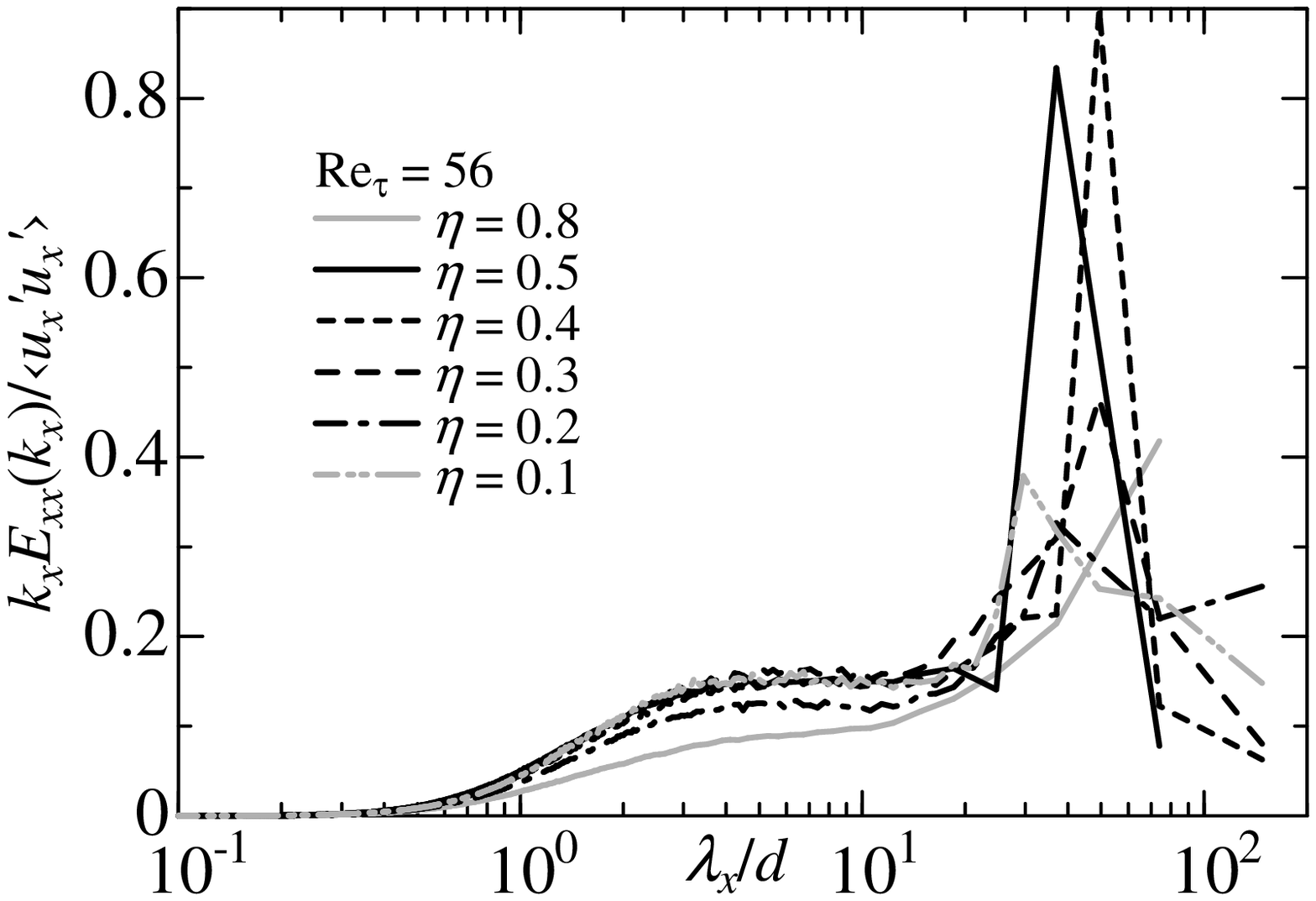} \hspace{1em}
(b) \includegraphics[width=70mm]{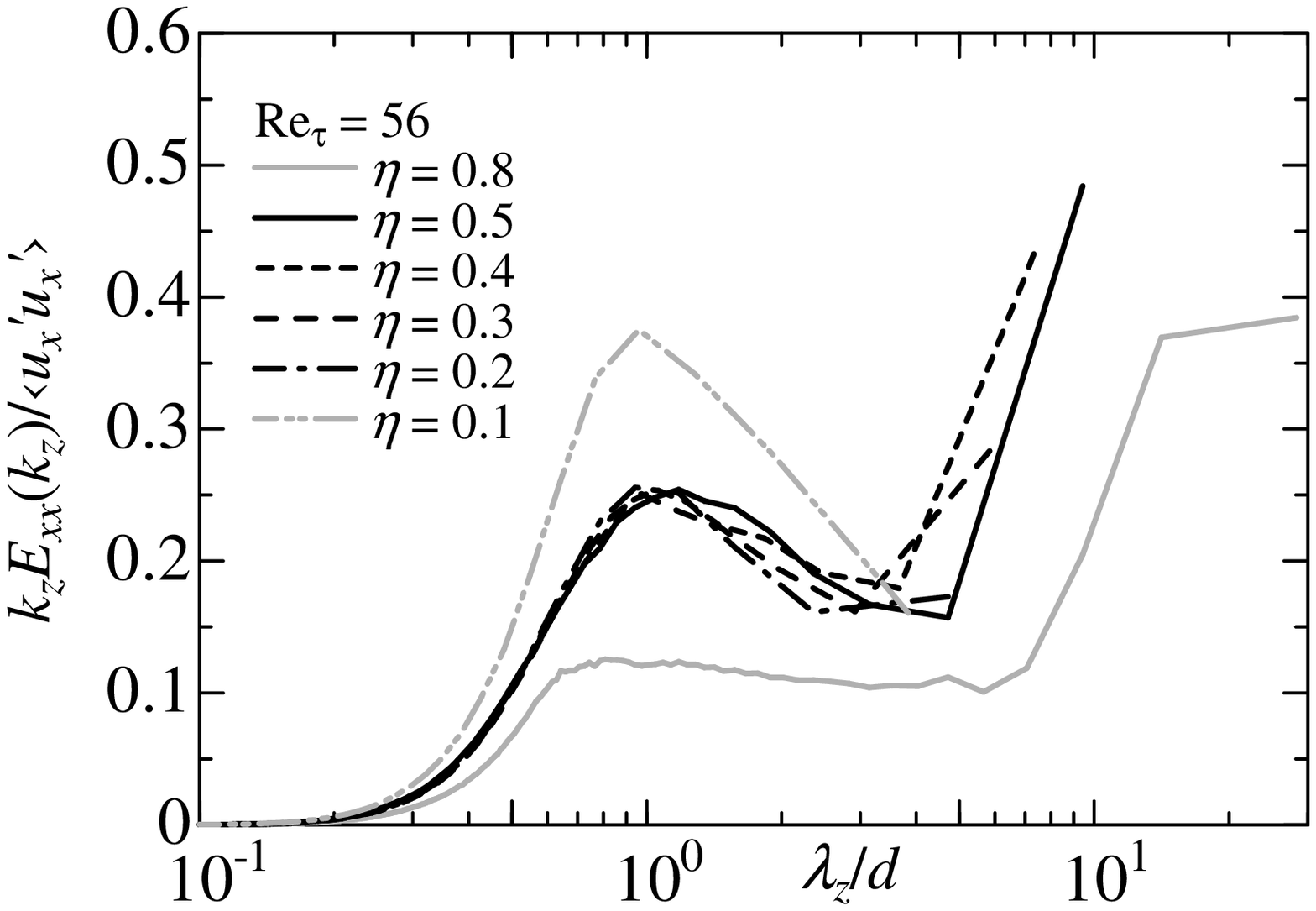}\\
(c) \includegraphics[width=70mm]{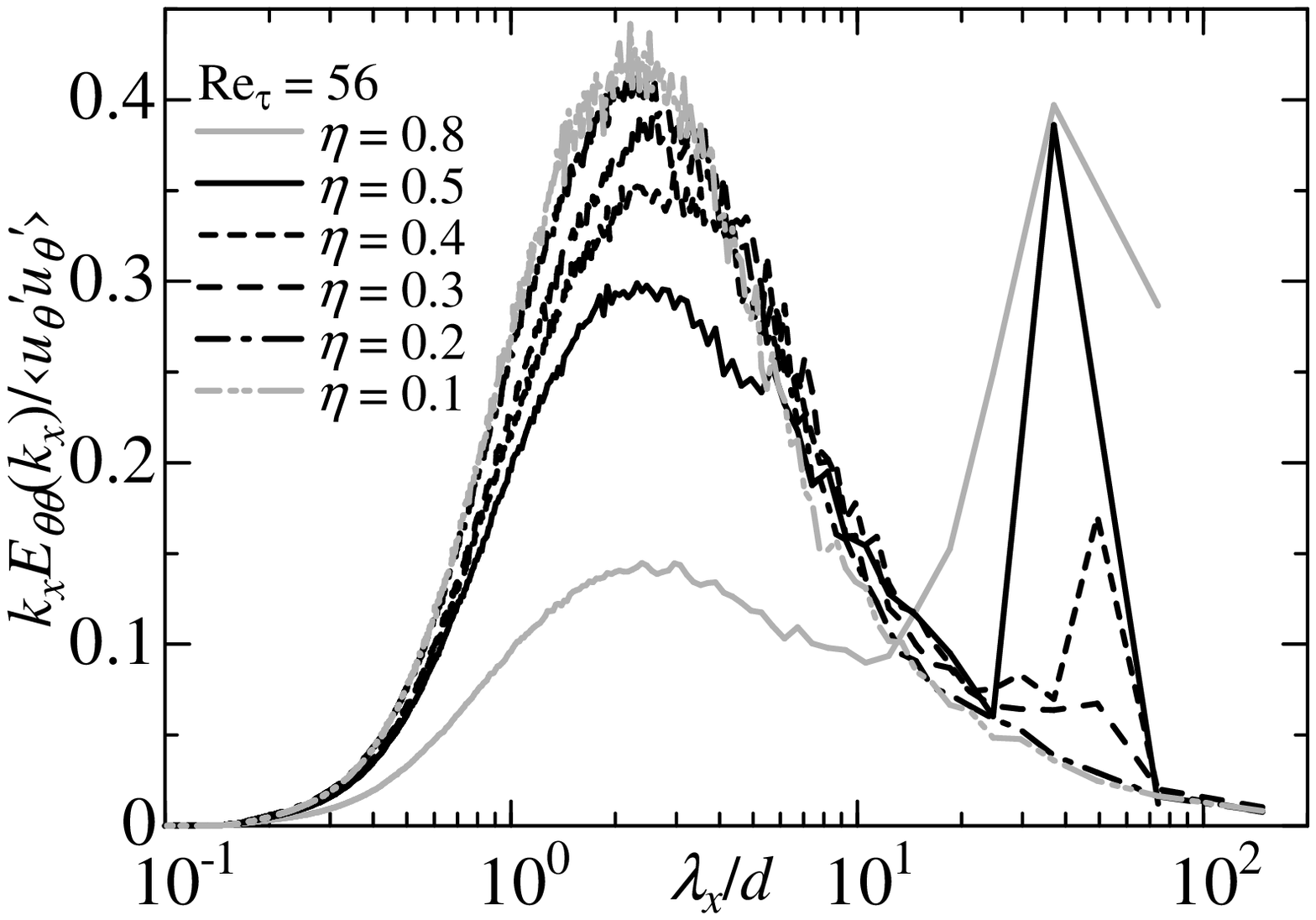} \hspace{1em}
(d) \includegraphics[width=70mm]{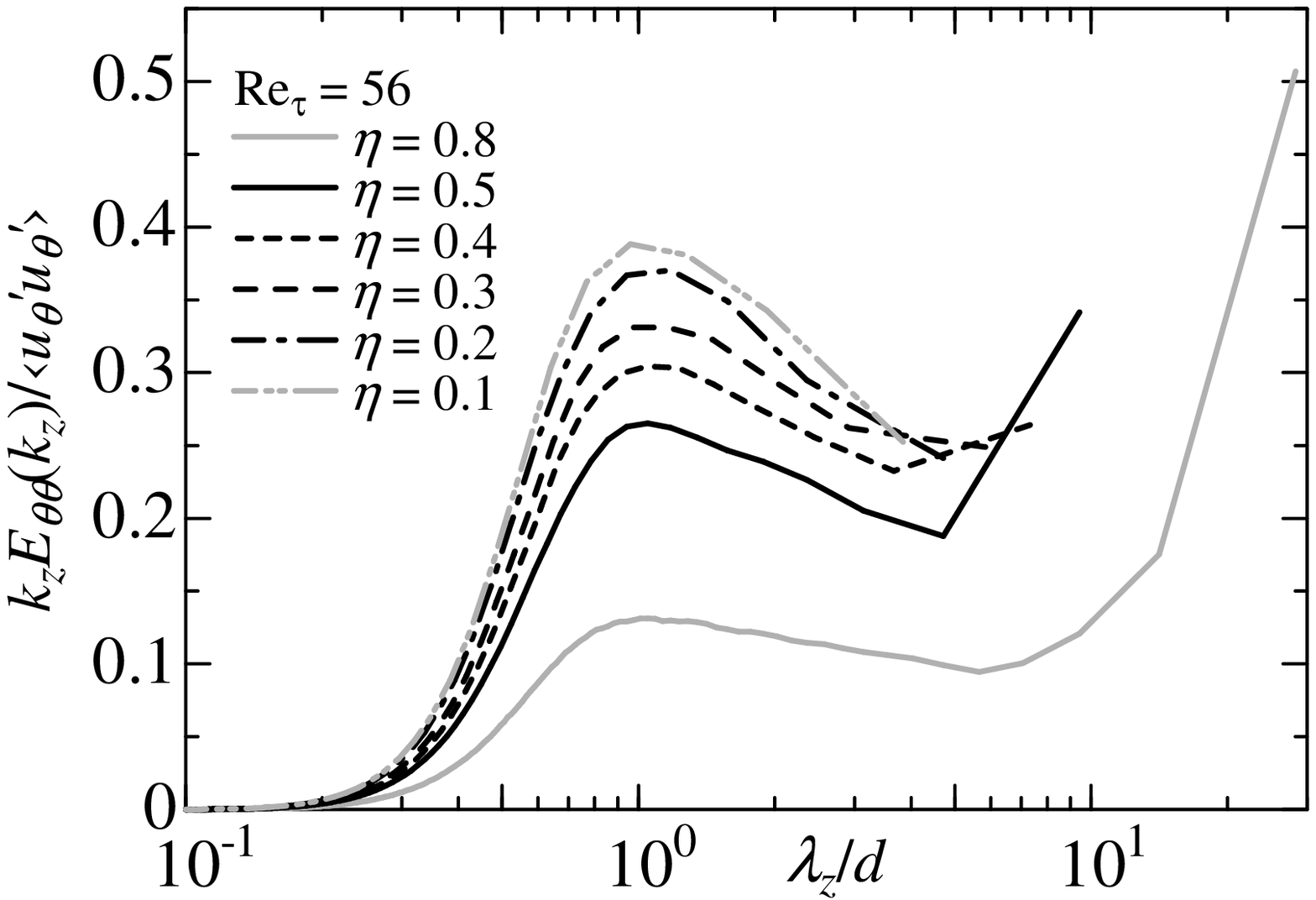}
\vspace{-1.5em}
\end{center}
\caption{Pre-multiplied energy spectra for $u'_x$ and $u'_\theta$ as a function of either wavelength $\lambda_x/d$ or $\lambda_z/d$ at mid-gap : (a) streamwise spectra of  $u'_x$, (b) spanwise spectra of  $u'_x$, (c) streamwise spectra of  $u'_\theta$, and (d) spanwise spectra of  $u'_\theta$. The streamwise and azimuthal wavenumbers are defined as $k_x=2\pi/\lambda_x$ and $k_z=2\pi/\lambda_z$, respectively. \red{The spectra} are normalized by an ensemble-averaged value of $u'_iu'_i$ at mid-gap.}
\label{fig:pmes}
\end{figure}

In order to evaluate which large-scale components are present here, time-averaged pre-multiplied energy spectra evaluated at gap center are shown in Fig.~\ref{fig:pmes} for ${\rm Re}_{\tau}=56$ (for which all flows are spatially intermittent) and for different values of $\eta$ from 0.1 to 0.8. \red{Energy spectra are based on the two-dimensional Fourier transform  $\widehat{u_i} (k_x, y, k_z)$ of $u'_i (x, y, z)$ with respect to $x$ and $z$, for all $y$. The \emph{streamwise} one-dimensional spectra associated with the streamwise and azimuthal velocity fluctuations  $u'_x$ and $u'_{\theta}$ are defined respectively by
\begin{equation}
E_{xx} (k_x,y) 
= \frac{L_x L_z}{(2\pi)^2 T} \int_t^{t+T} \int_0^\infty \Re \left( \widehat{u_x} \widehat{u_x}^* \right) {\rm d} k_z {\rm d}t,~
E_{\theta \theta} (k_x,y) 
= \frac{L_x L_z}{(2\pi)^2 T} \int_t^{t+T} \int_0^\infty \Re \left( \widehat{u_\theta } \widehat{u_\theta }^* \right) {\rm d} k_z {\rm d}t.
\end{equation}
The asterisk represents complex conjugation, $T$ is the averaging time, and $\Re$ denotes the real part.
The proportionality constant is chosen so that 
\begin{equation}
\int E_{xx} (k_x) {\rm d} k_x = \int E_{xx} (k_z) {\rm d} k_z = \left<u_x'u_x' \right>, ~~
\int E_{\theta \theta} (k_x) {\rm d} k_x = \int E_{\theta \theta} (k_z) {\rm d} k_z = \left< u_\theta'u_\theta' \right>
\end{equation}
as function of either $k_x$ or $k_\theta$, where averaging $\left< \cdot \right>$ is performed over $x$, $\theta$, and $t$.}

Careful analysis of Fig.~\ref{fig:pmes} reveals robust features. Small scales associated with turbulent fluctuations are present around $\lambda_x/d \approx 2$--3 and $\lambda_z/d \approx 1$ in all directions for all components. The situation is different for large-scale velocity components : out of the four figures in Fig.~\ref{fig:pmes}, only $u_x$ as a function of $x$ displays large scales for all value of $\eta$. These large scales do not appear strongly separated from the smaller ones, and are located at $\lambda_x/d \approx 15$--70. 
Neither $u_{\theta}$ as a function of either $x$ or $\theta$, nor $u_x$ as a function of $\theta$, possesses such a robust large-scale component. Only as $\eta$ exceeds $0.3$, do well-separated peaks at similar large scales emerge in each spectrum. Important \red{information} can be deduced from these spectra. First, should \red{large scales} be found, the corresponding cut-off can be located safely in the intervals $\lambda_x^c/d \approx 10$--20 and $\lambda_z^c/d \approx 2$--5. Secondly, by construction the wavelengths in the spectra are limited by the box dimensions $L_x$ and $L_z$. A clear difference emerges between Figs.~\ref{fig:pmes}(a, c) on one hand (showing the $\lambda_x$ dependence), and Figs.~\ref{fig:pmes}(b, d) on the other hand (showing the $\lambda_z$ dependence) : in the latter case the occurrence of an azimuthal large-scale peak for both $x$ and $\theta$ components is ruled out when $L_z \ll \lambda_z^c$. In other words, the spanwise large-scale component is present only for sufficient azimuthal extent, whereas streamwise large-scale modulations are always present as long as the flow features spatial intermittency. We emphasize here that the azimuthal extent should be measured in units of $d=R_o-R_i$, since streaks and streamwise vortices (forming the small scales) scale with the gap size $d$ rather than with any of the two radii $R_o$ or $R_i$. Wall units are another candidate, but large-scale structures of interest are generally considered to scale with outer units. Our hypothesis here is that the occurrence of azimuthal large-scale flows depends directly on the azimuthal extent $L_z/d$, itself a function of $\eta$.\\

\begin{figure}[t]
\begin{center}
\includegraphics[width=90mm]{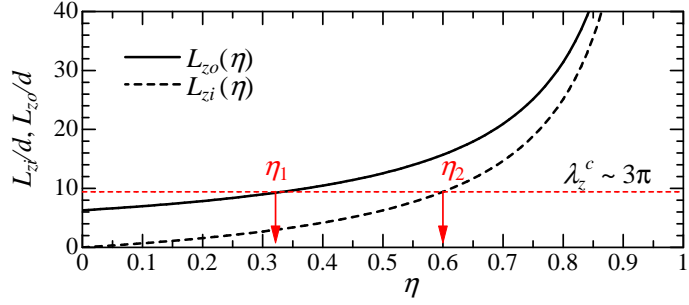}
\vspace{-1.5em}
\end{center}
\caption{Azimuthal length as a function of radius ratio.}
\label{fig:lz}
\end{figure}

The quantity $L_z$, however, depends on the value of $r$. It is simpler to focus on the inner and outer azimuthal extents $L_{zi}$ and $L_{zo}$, given respectively by
\begin{eqnarray}
L_{zi}=2\pi d\frac{\eta}{1-\eta}\\
L_{zo}=2\pi d\frac{1}{1-\eta}.
\label{eq:Lz}
\end{eqnarray}
Both quantities (expressed in units of $d$) are plotted as functions of the radius ratio $\eta$ in Fig.~\ref{fig:lz}.

Let us fix rather arbitrarily a cut-off value $\lambda_{z}^c$. The intersection of the horizontal line $L_z=\lambda_z^c$ with the curves $L_{zo}(\eta)$ and $L_{zi}(\eta)$ in Fig. \ref{fig:lz} defines two values of $\eta$, respectively $\eta_1$ and $\eta_2$. This leads to three distinct ranges of values for $\eta$ : 
\begin{itemize}
\item for $0<\eta \le \eta_1$, there is no space for large scales in the azimuthal direction, neither at the inner not at the outer wall. As a consequence $\overline{U_\theta}=0$ for all $r$, and $\partial_x \overline{U_x}=0$.
\item for $\eta_2 \le \eta \le 1$, azimuthal large scales can form at both inner and outer walls : $\partial_{\theta} \overline{U_\theta} = - \partial_x \overline{U_x}$ with $\overline{U_\theta} \neq 0$. The situation is then analogous to the planar case.
\item for $\eta_1 \le \eta \le \eta_2$, the situation is mixed : azimuthal large scale flows cannot be accommodated at all locations in the cross-section. A probabilistic approach is required.
\end{itemize}
For instance, choosing $\lambda_z^c/d=3\pi$ leads to $\eta_1 = 1/3$ and $\eta_2 = 3/5$, which is consistent with our observations.
While the classification above is not useful in practice to predict accurately the transition thresholds $\eta_1$ and $\eta_2$ (mainly because of the difficulty to define a unique cut-off value $\lambda_z^c({\rm Re}_{\tau})$), it captures the main physical idea : the presence of confinement in the azimuthal direction defines the two extreme regimes of straight interfaces (associated with puffs) or oblique interfaces (associated with oblique stripes). \red{In addition,} there is a range of value of $\eta$ for which there is a \red{probability of observing} both types of interfaces (and hence both puffs and stripes) in the same flow at different times and/or different positions.\\

\begin{figure}[t]
\begin{center}
(a)\includegraphics[width=50mm]{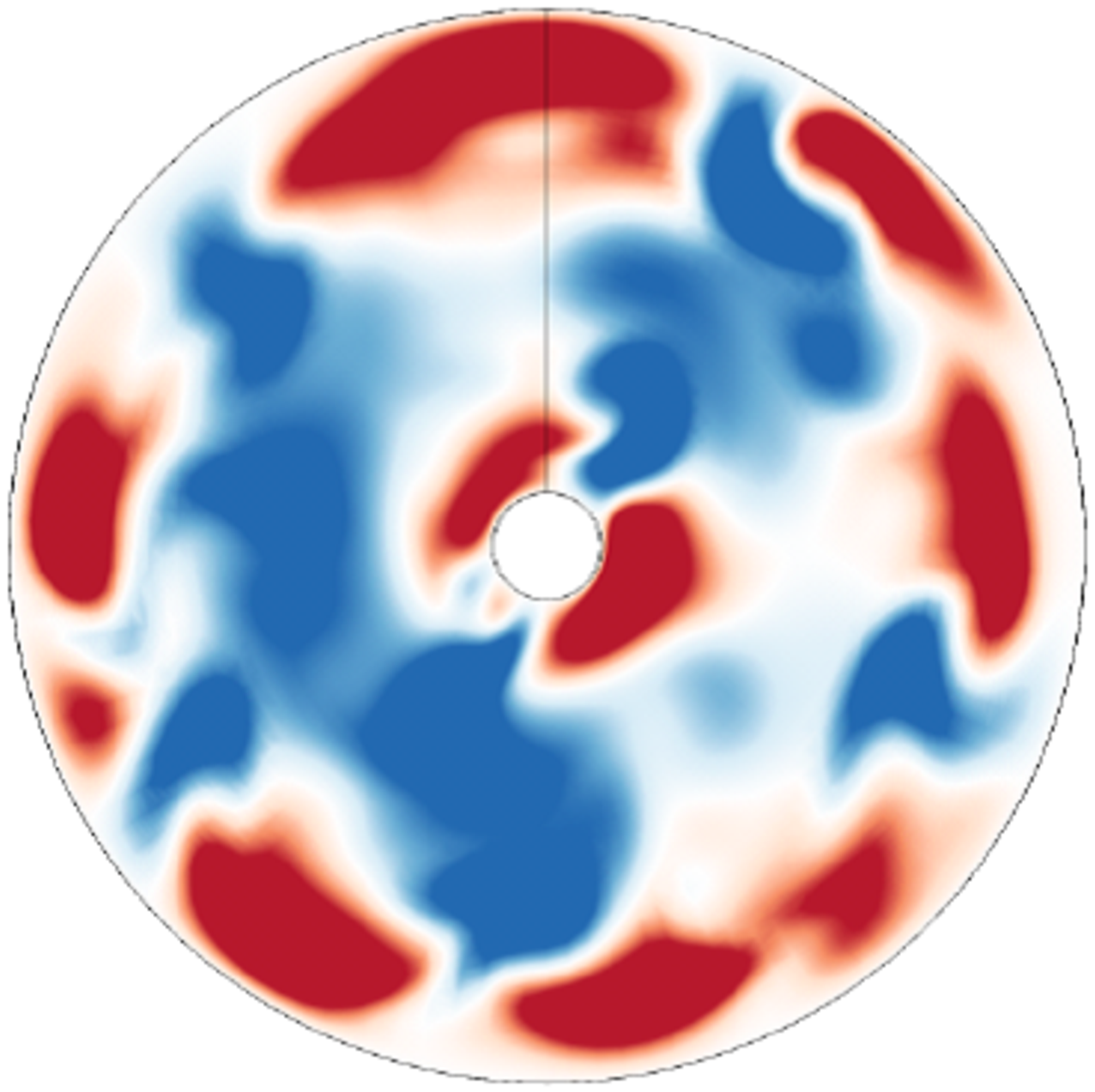}
(b)\includegraphics[width=50mm]{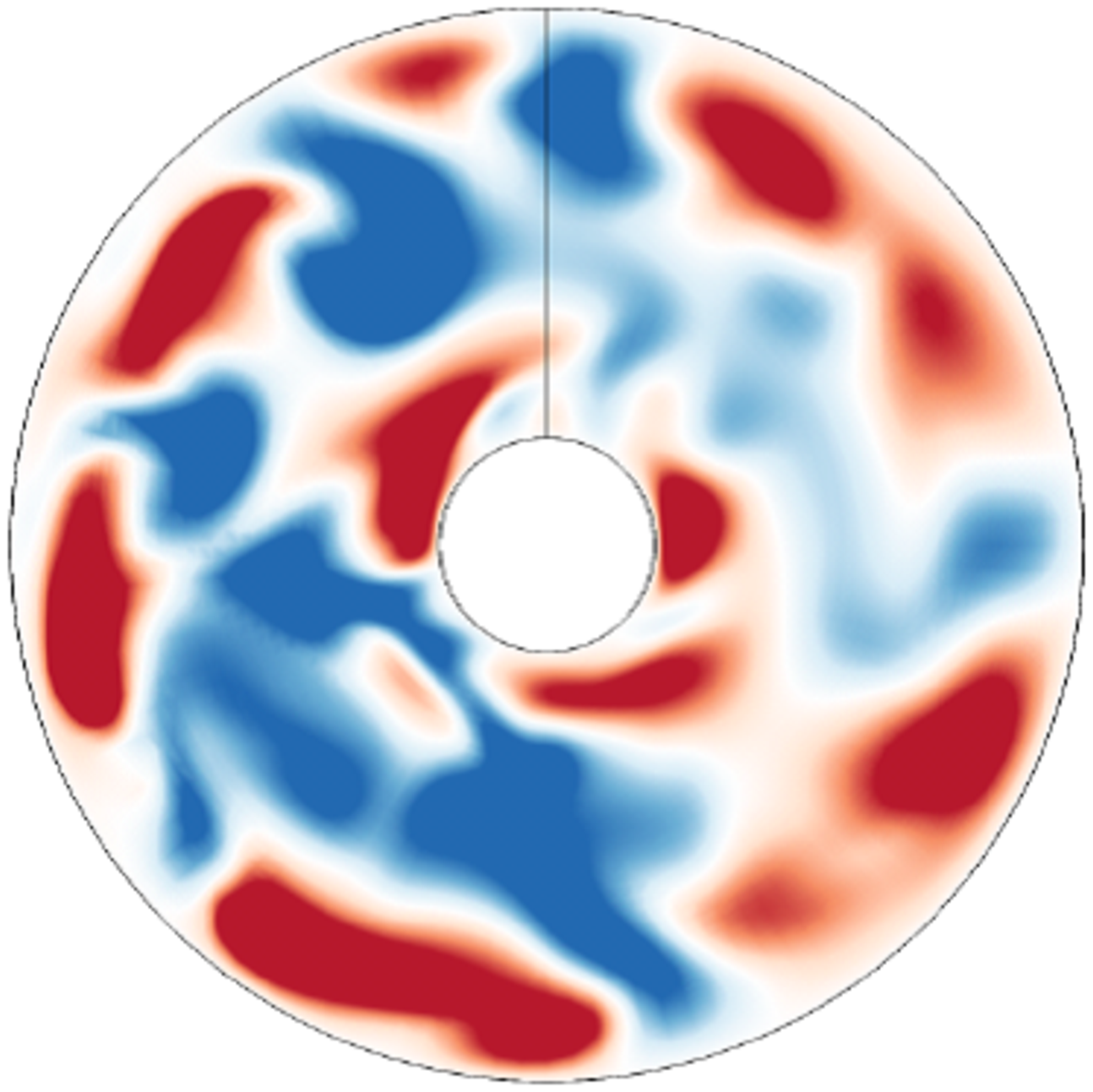}
(c)\includegraphics[width=50mm]{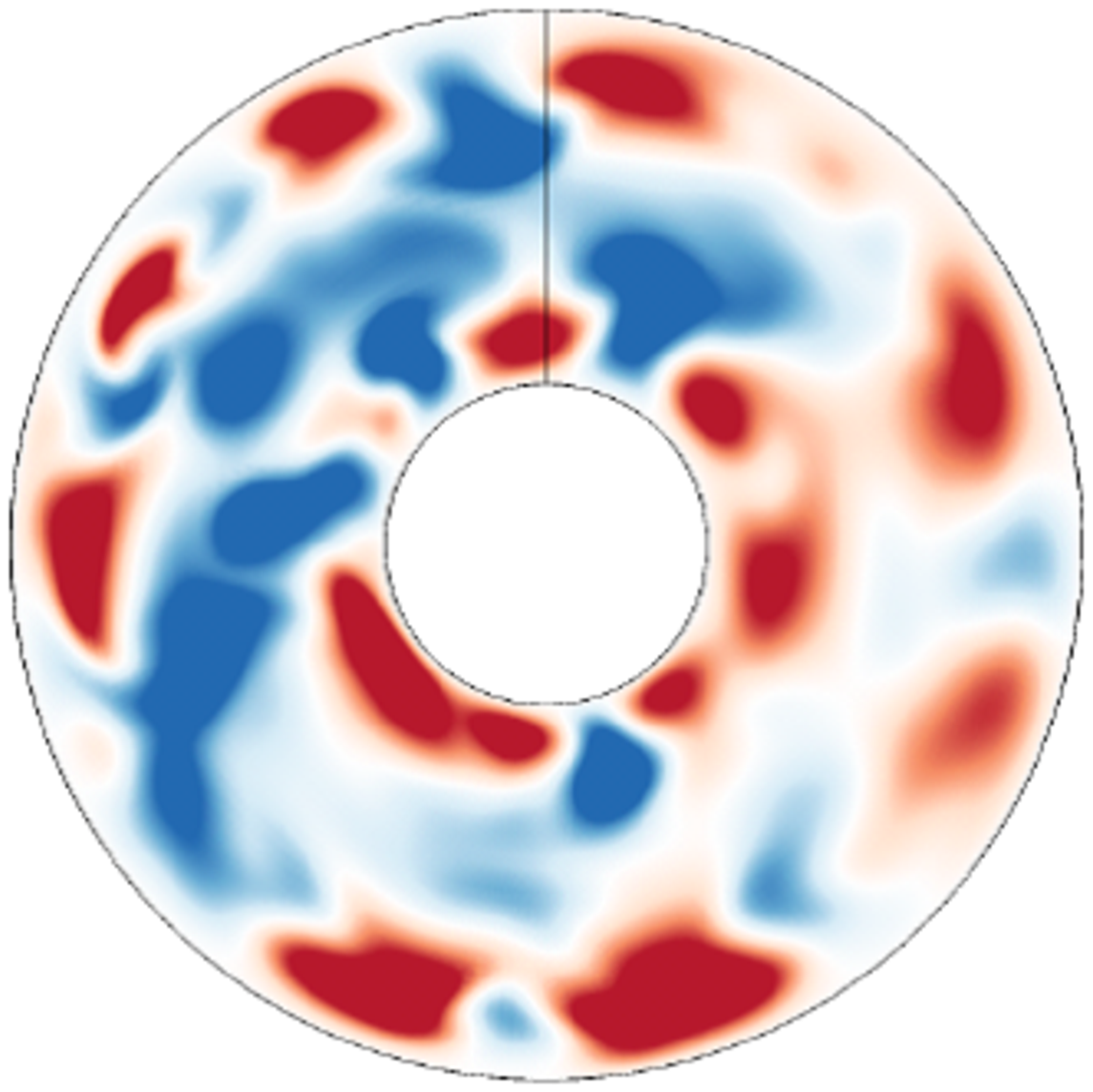}\\ \vspace{1em}
(d)\includegraphics[width=50mm]{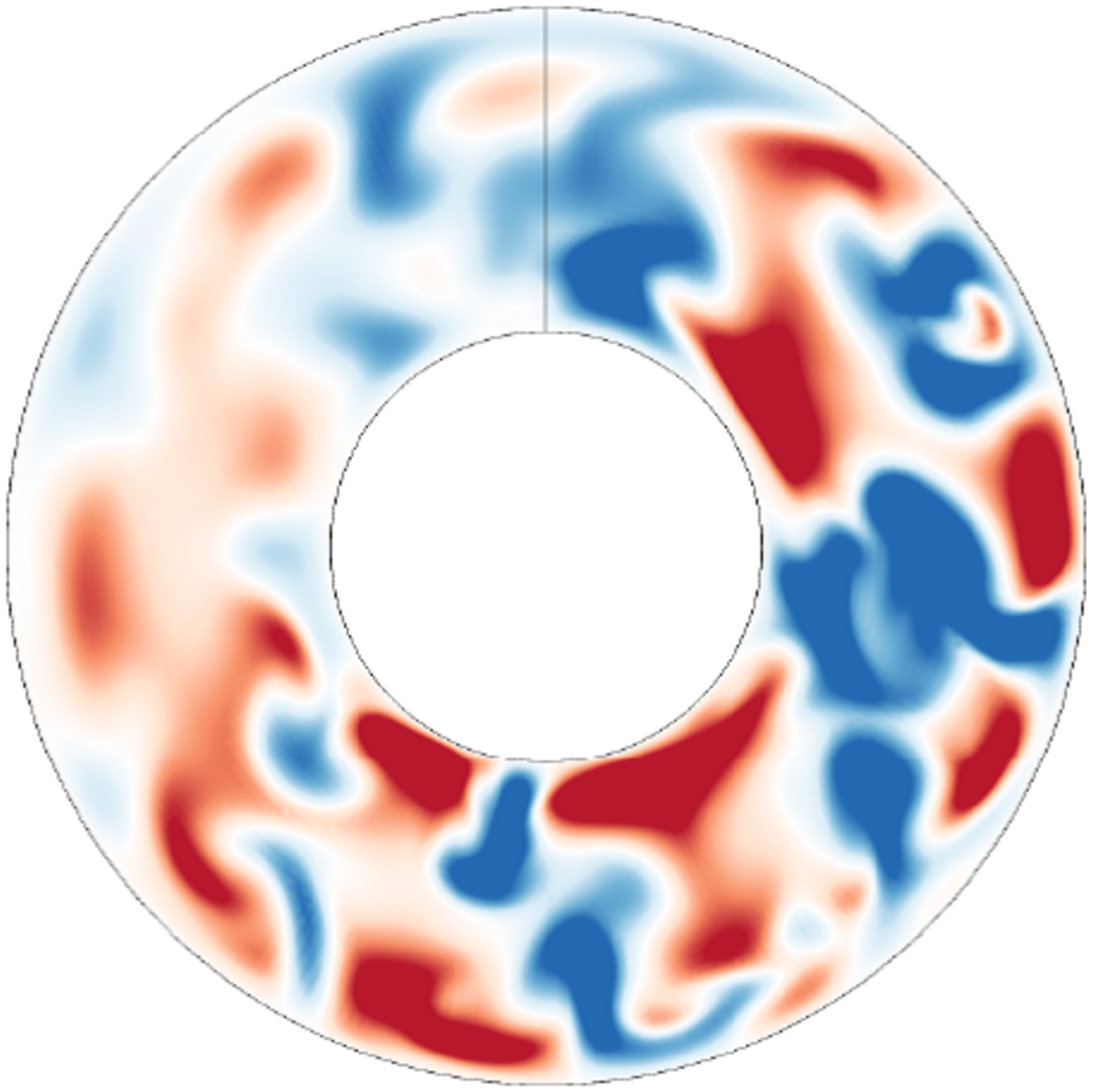}
(e)\includegraphics[width=50mm]{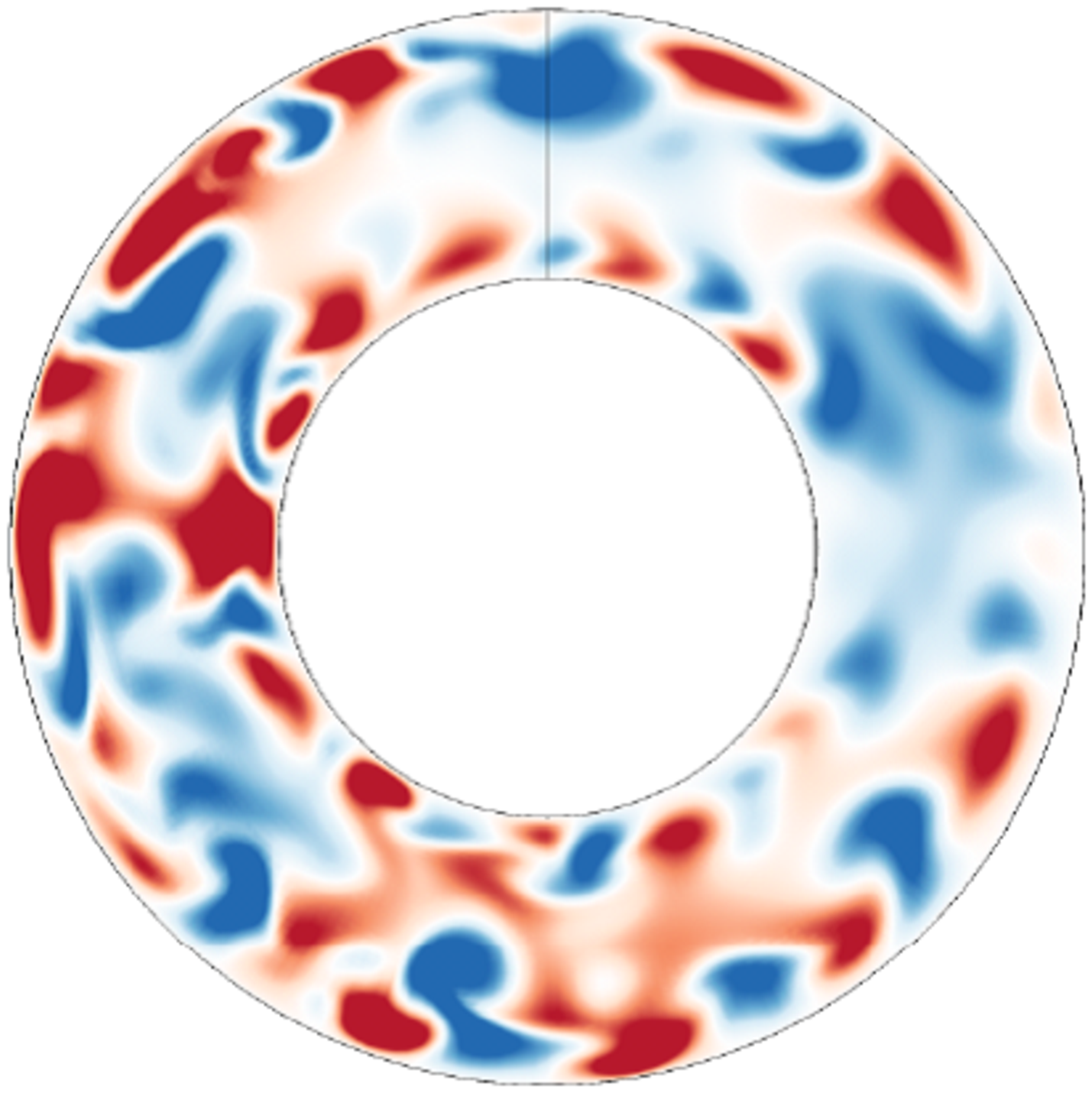}
(f)\includegraphics[width=50mm]{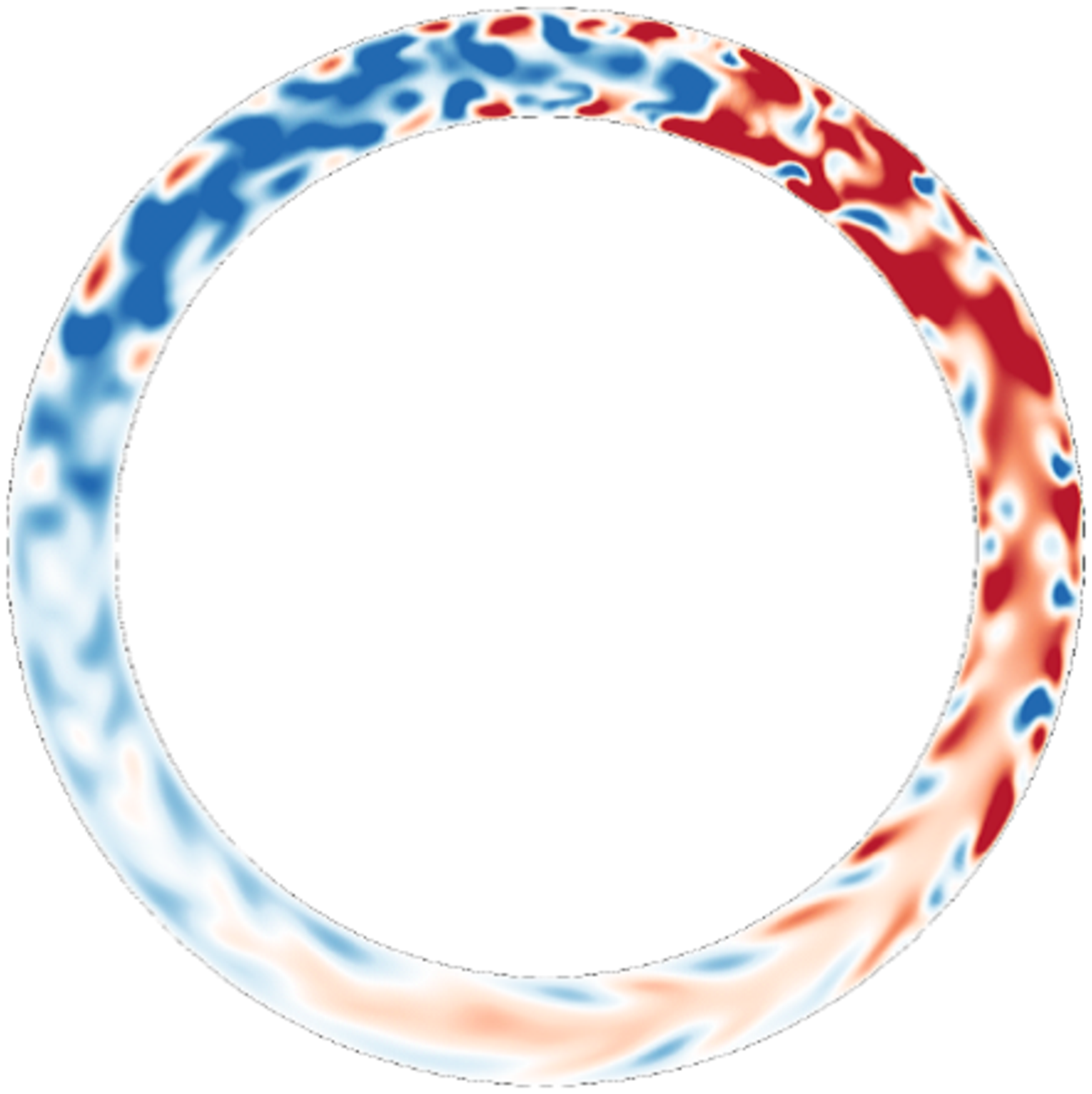}
\vspace{-0.5em}
\end{center}
\caption{Two-dimensional cross-sections ($r, \theta$) of instantaneous $u'_x$ for $\eta$ = (a) 0.1, (b) 0.2, (c) 0.3, (d) 0.4, (e) 0.5, and (f) 0.8 at Re$_\tau=52$. Colourmap from $-3.0u_{\tau}$ (blue) to $+3.0u_{\tau}$ (red).}
\label{fig:rtu}
\end{figure}

\begin{figure}[t]
\begin{center}
(a)\includegraphics[width=50mm]{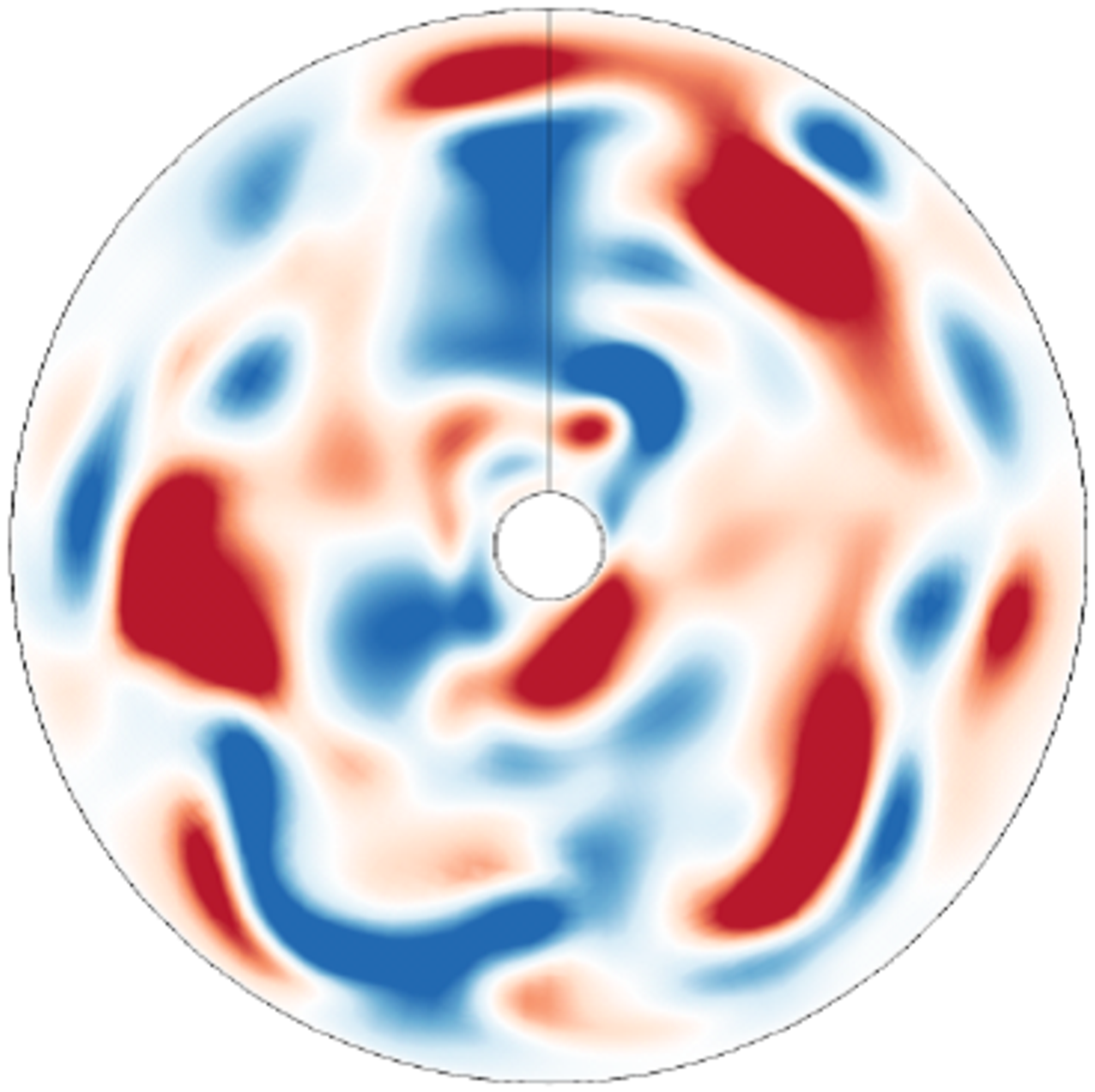}
(b)\includegraphics[width=50mm]{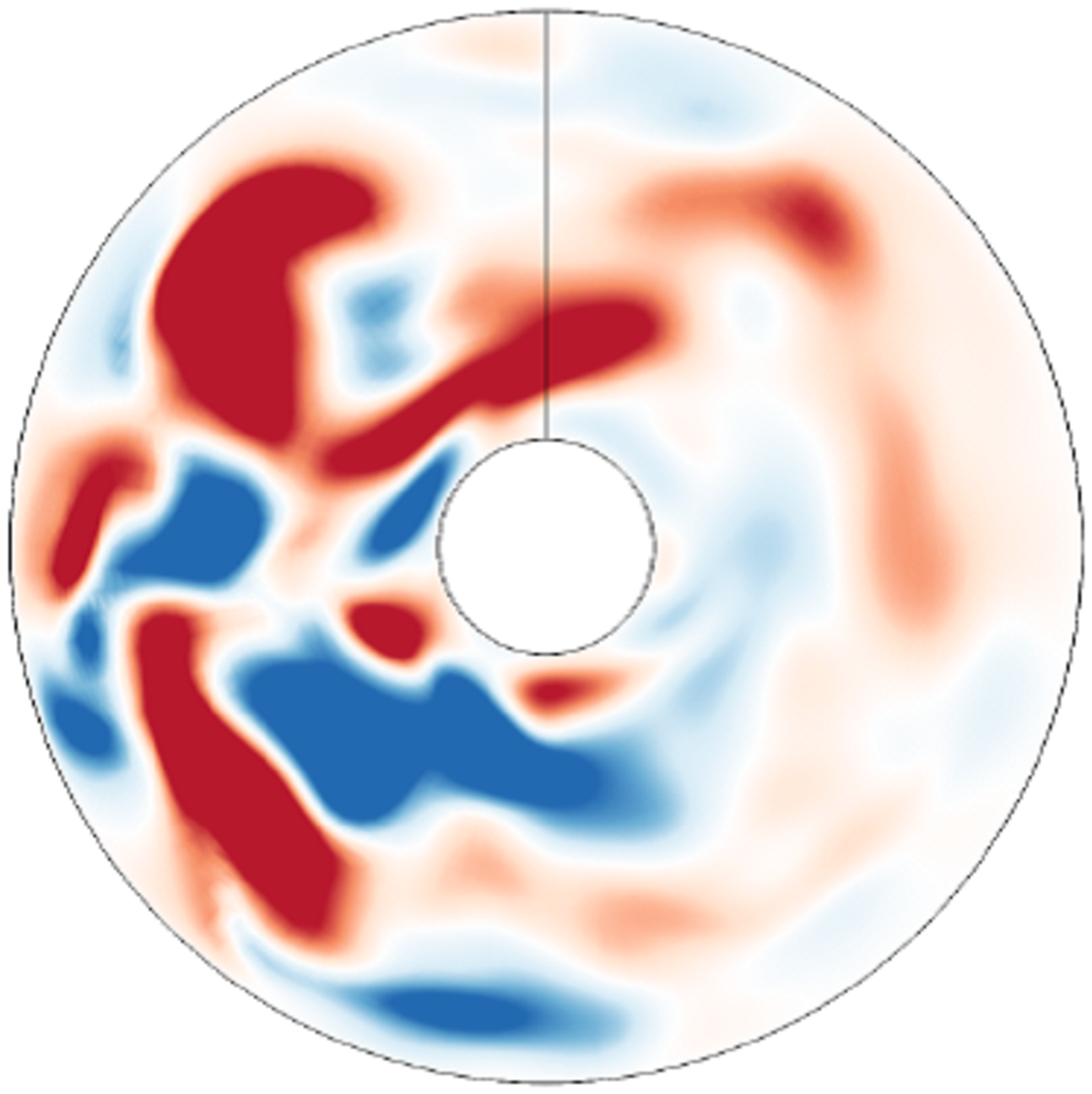}
(c)\includegraphics[width=50mm]{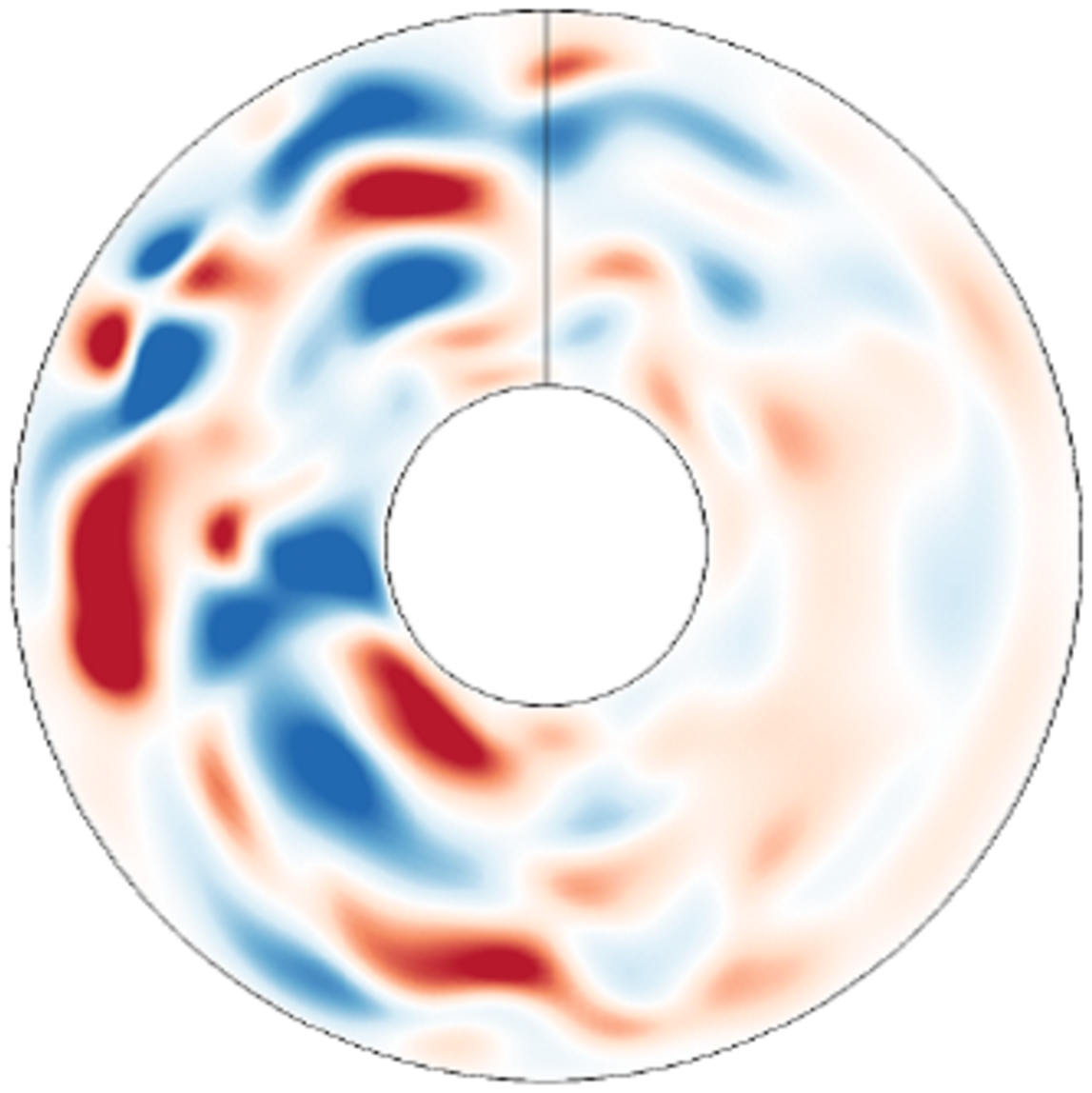}\\ \vspace{1em}
(d)\includegraphics[width=50mm]{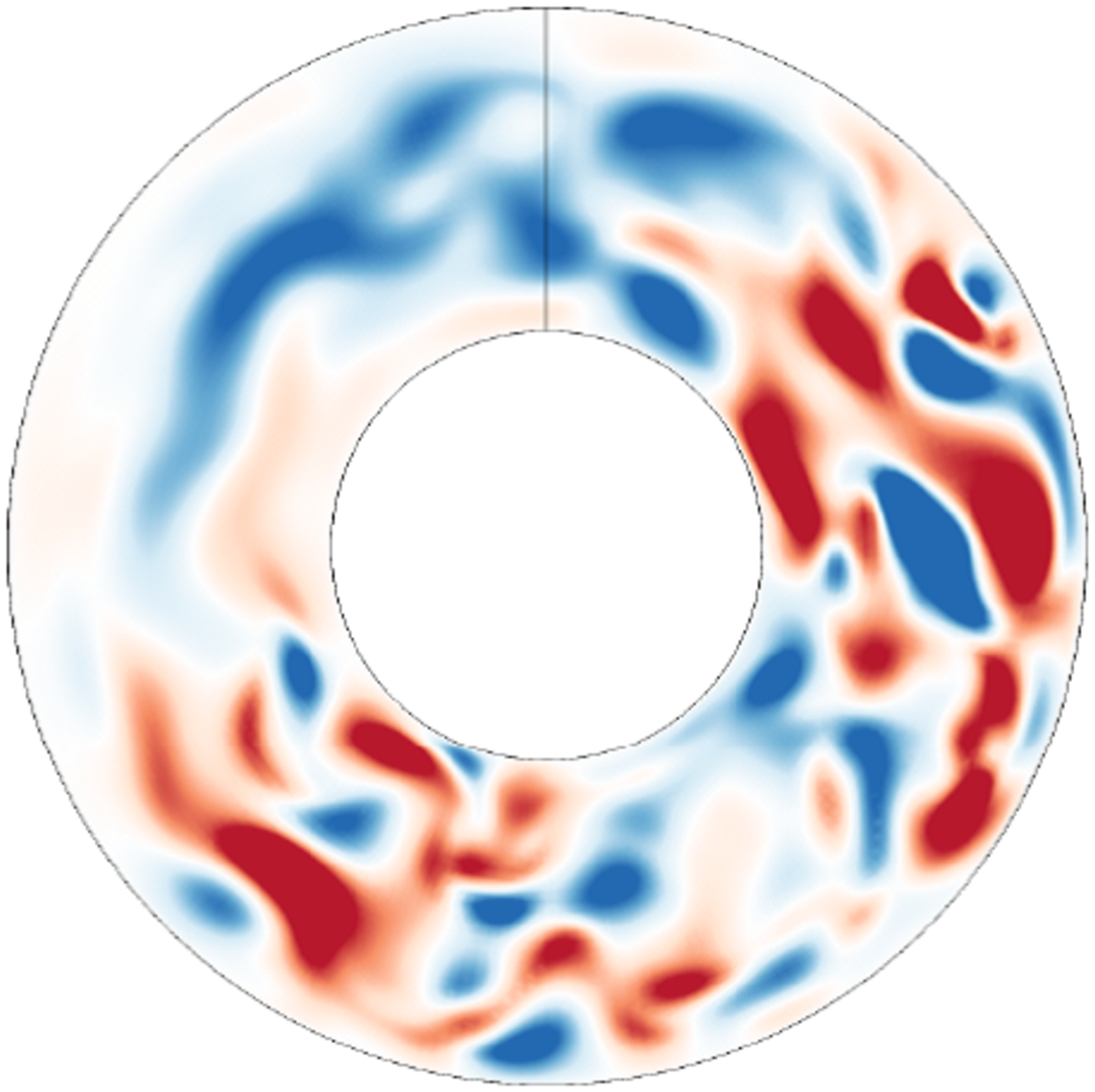}
(e)\includegraphics[width=50mm]{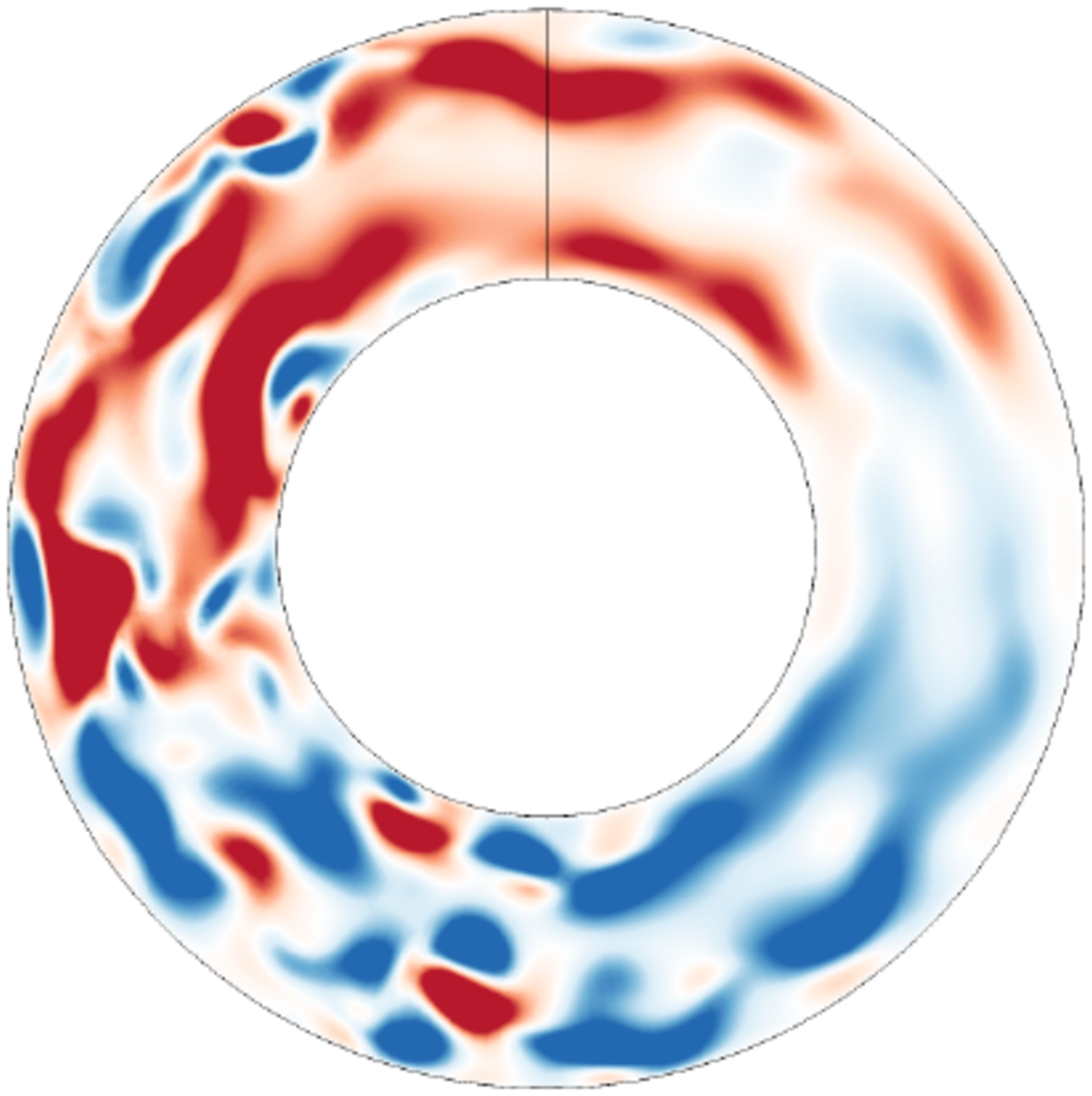}
(f)\includegraphics[width=50mm]{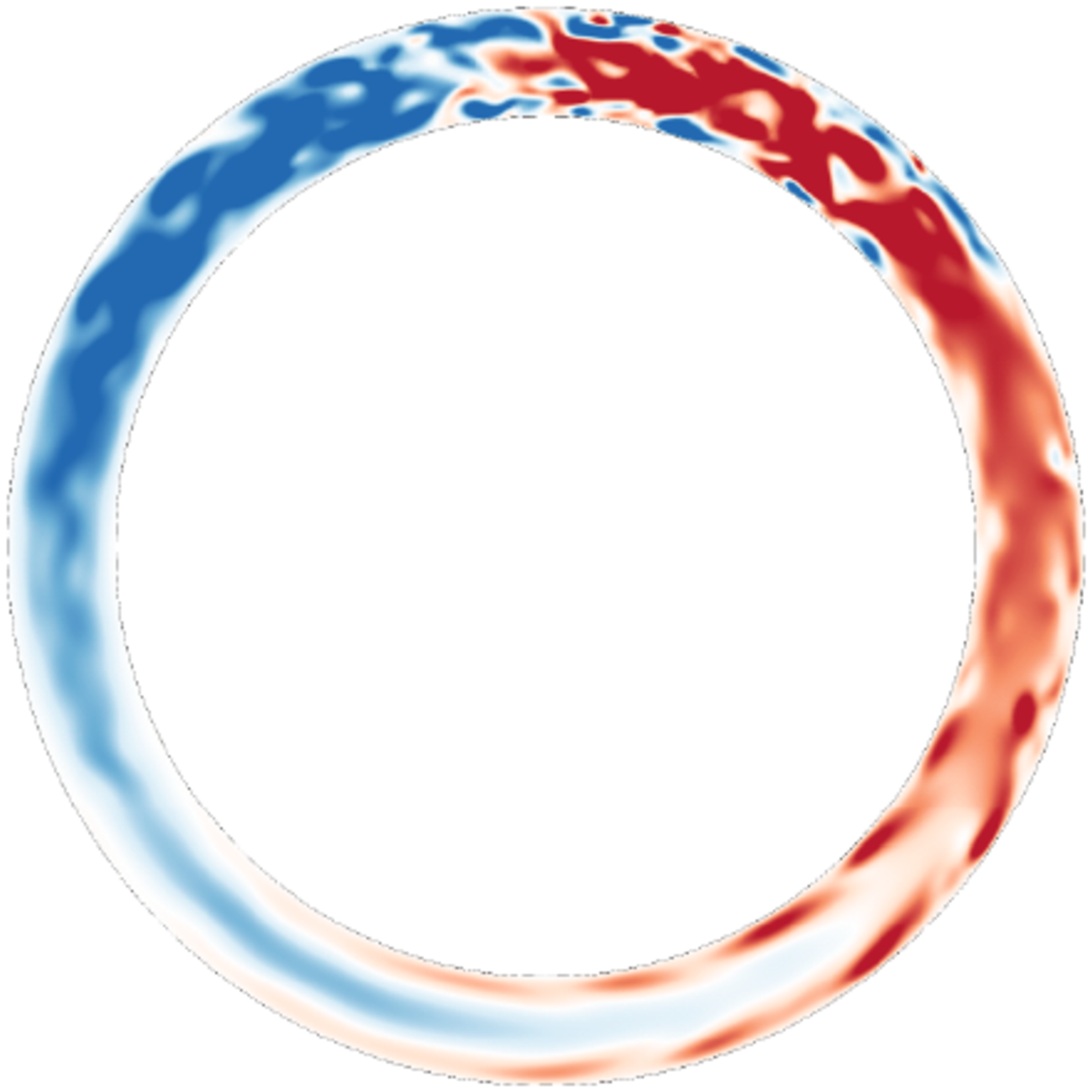}
\vspace{-0.5em}
\end{center}
\caption{Two-dimensional cross-sections ($r, \theta$) of instantaneous $ru'_\theta$ for $\eta$ = (a) 0.1, (b) 0.2, (c) 0.3, (d) 0.4, (e) 0.5, and (f) 0.8 at Re$_\tau=52$. Colourmap from $-1.0u_{\tau}$ (blue) to $+1.0u_{\tau}$ (red): the positive direction of $\theta$ is clockwise in the plots.}
\label{fig:rtw}
\end{figure}

It can be useful to investigate cross-sections of the flow in the different regime to understand the implications of the previous hypothesis.
The two-dimensional contours of $u'_x$ and $u'_\theta$ in arbitrary chosen ($r$-$\theta$) cross-sections are shown in Figs.~\ref{fig:rtu} and \ref{fig:rtw}, respectively, for different values of $\eta$.

The isocontours of $u'_x$ in Fig.~\ref{fig:rtu} allow one to count the numbers of streaks present in the vicinity of each wall. Each of these streaks has a radial and azimuthal extent $\approx d/2$, and the analysis of the spectra in Fig. \ref{fig:pmes} also suggests that their streamwise extent is approximatively $2d$. In principle one would expect the streak size to scale in inner units $\nu/u_{\tau}$, however the range of values of ${\rm Re}_{\tau}$ investigated here is relatively narrow and we prefer to report the dimensions of the streaks in (outer) units of $d$, as the relation $\lambda_z \approx d$ emphasizes their quasi-circular cross-section.
\red{In the case of the straight puffs found for $\eta = 0.1$, $u'_x$ shows no large-scale azimuthal modulation, while high-order modulations (streaks) can be easily noted.} With increasing $\eta$, the number of streaks increases as the azimuthal extent increases, a supplementary confirmation that the relevant lengthscale here is $d$ rather than $R_o$ or $R_i$. For $\eta = 0.5$ and 0.8 (Fig. \ref{fig:rtu}(d, e)), the low-order non-axisymmetric modulation of $u'_x$ is the direct signature of the helix-shaped turbulence. To a lesser degree, such a modulation can also be visually detected for $\eta = 0.2$ and 0.3 (Fig.~\ref{fig:rtu}(b, c)). Similar conclusions can be drawn from the isocontours of $u'_{\theta}$ in Fig.~\ref{fig:rtw} as well.\\

\subsection{4.3 Statistics of transverse large-scale flows}

This section is now devoted to a quantitative investigation of the orientation of the large-scale flow near the interfaces for varying radius ratio $\eta$. We begin by describing in more detail how data from the previous direct numerical simulations is post-processed. Based on the apparent scale separation in the spectra from Fig.~\ref{fig:pmes}, we consider a low-pass filter $\mathcal{L}$ whose kernel in spectral $(k_x,k_z)$ space reads
\begin{eqnarray}
\left|k_x \right| \le \frac{2\pi}{20d},~ \left|k_z \right| \le \frac{2\pi}{2d}
\label{eq:kernel}
\end{eqnarray}
The original velocity fields ${\bf u}(x,r,\theta,t)$ are transformed via $\mathcal{L}$ into filtered fields ${\bf U}(x,r,\theta,t)$. The two-dimensional wall-integrated large-scale flow $(\overline{U_x}, \overline{U}_{\theta})$ is then computed using the definitions in Eq.~(\ref{def_int}).\\

\begin{figure}[t]
\begin{center}
\includegraphics[width=85mm]{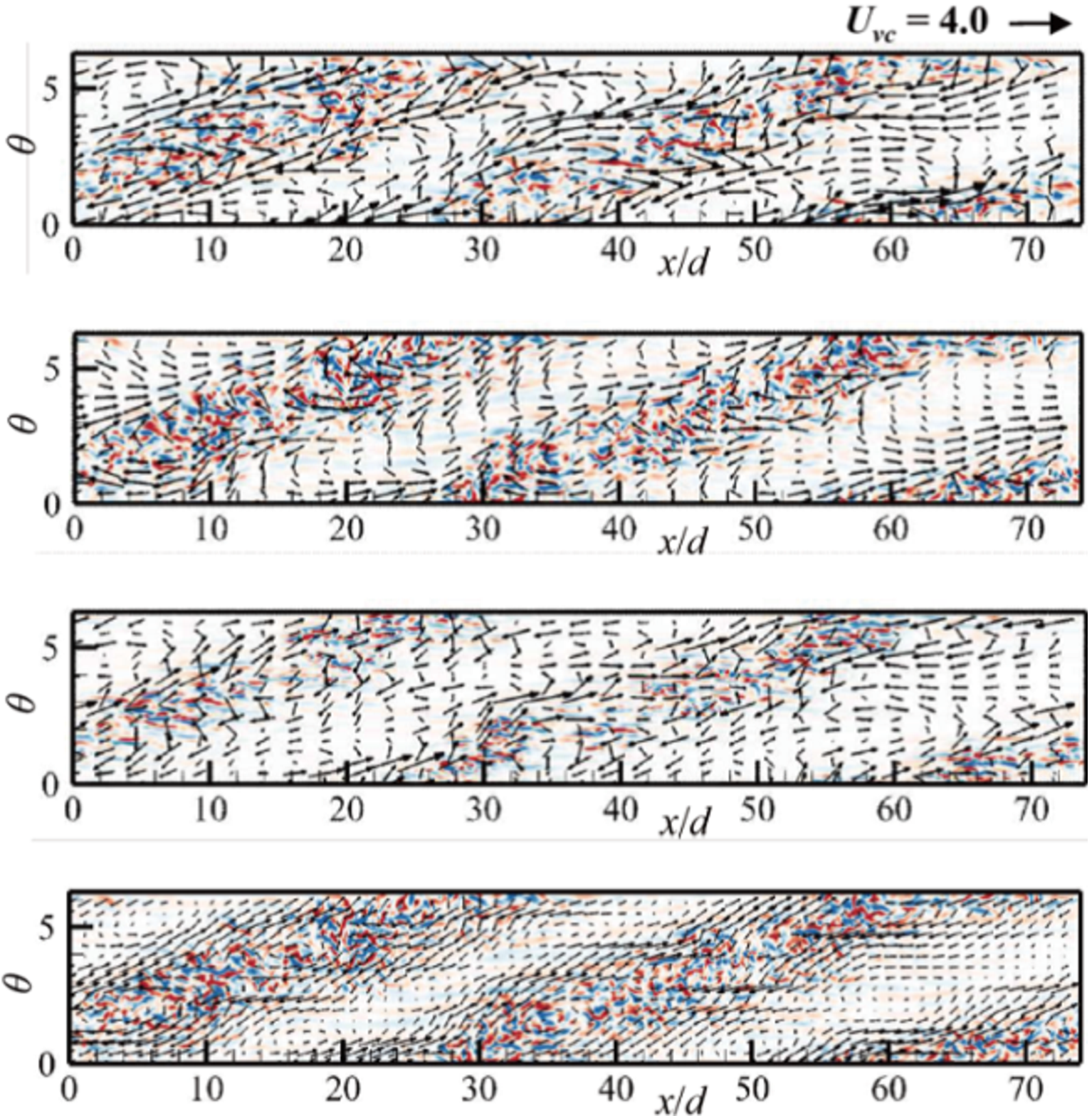}\\
\vspace{-23em} \hspace{-31em}(a)\\
\vspace{  6em} \hspace{-31em}(b)\\
\vspace{  6em} \hspace{-31em}(c)\\
\vspace{  6em} \hspace{-31em}(d)
\vspace{-0.5em}
\end{center}
\caption{Two-dimensional contours of instantaneous $u'_r$ (blue, red) = ($-1.0, 1.0$) and large-scale flow ($U_x, rU_\theta$) \red{as vectors} at $y/d$ = (a) 0.19, (b) 0.5, and (c) 0.81, for $\eta = 0.5$ and Re$_\tau=56$. \red{Wall-normal-averaged} large-scale flow ($\overline{U_x}, \overline{U_\theta}$) with contours of $u'_r$ at mid-gap is shown in (d). The reference-vector length $U_\textit{VC}$ is shown in the top right of (a).}
\label{fig:2d05_lsf}

\vspace{1.0em}

\begin{center}
\includegraphics[width=85mm]{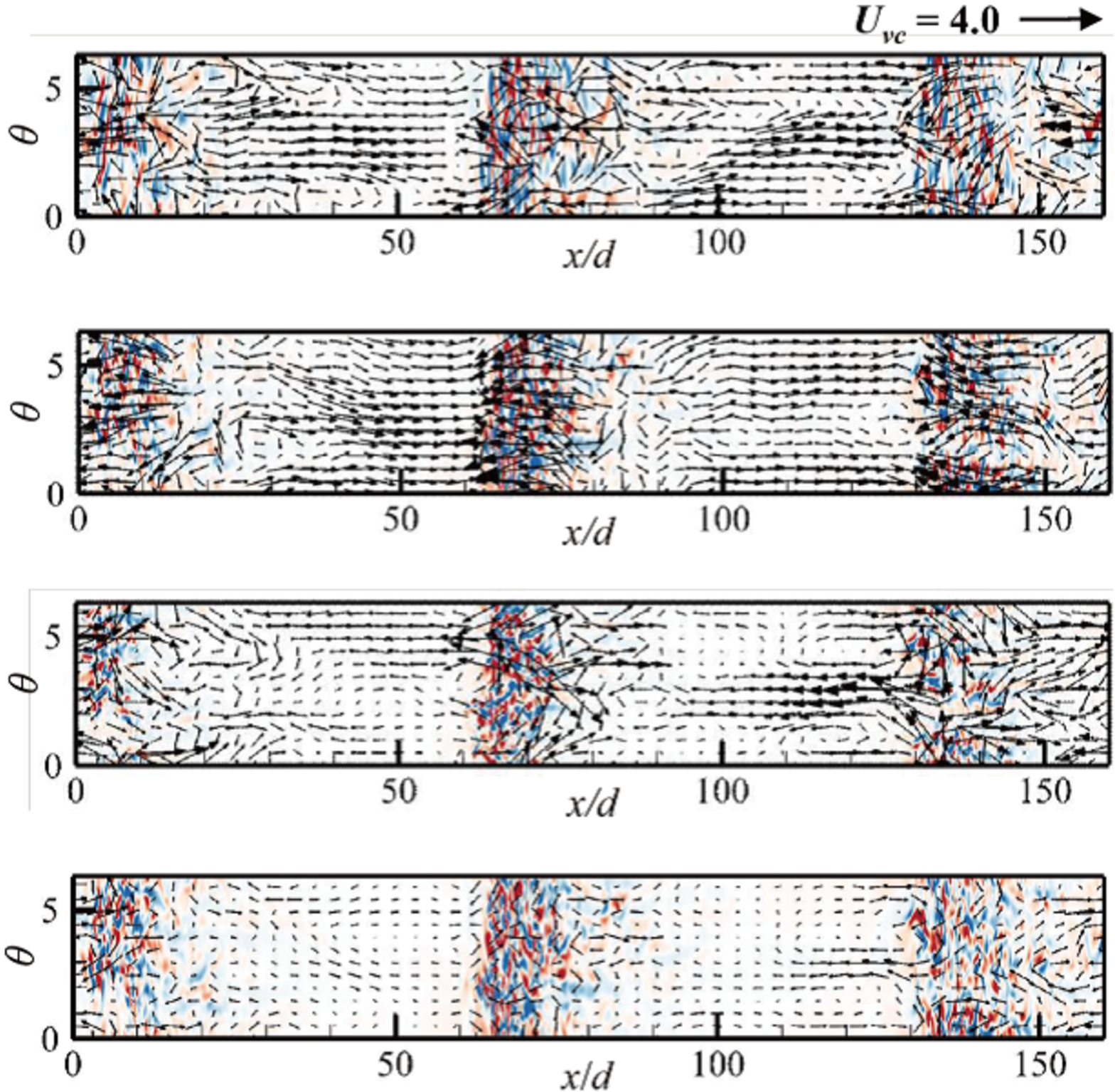}\\
\vspace{-23em} \hspace{-31em}(a)\\
\vspace{  6em} \hspace{-31em}(b)\\
\vspace{  6em} \hspace{-31em}(c)\\
\vspace{  6em} \hspace{-31em}(d)
\vspace{-0.5em}
\end{center}
\caption{Same as Fig.~\ref{fig:2d05_lsf}, but at $y/d$ = (a) 0.25, (b) 0.5, and (c) 0.75, for $\eta = 0.1$ and Re$_\tau=56$.}
\label{fig:2d01_lsf}
\end{figure}

An instantaneous snapshot of the large-scale flow ${\bf U}(x,r,\theta,t)$ is plotted for several values of $r$ together with its wall-integrated counterpart, namely, $\overline{U_x}(x,\theta,t)$ and $\overline{U_\theta}(x,\theta,t)$. The cases $\eta=0.5$ and $\eta=0.1$ are respectively shown in Figs.~\ref{fig:2d05_lsf} and \ref{fig:2d01_lsf}. These plots highlight the genuinely three-dimensional structure of the velocity field ${\bf U}$. For $\eta=0.5$, where turbulence clearly takes a helical shape, it can be verified that the local orientation of ${\bf U}$ does not necessarily match that of the laminar-turbulent interface. The two-dimensional counterpart $(\overline{U_x},\overline{U_\theta})$ however does point in a direction parallel to the interface, which confirms the previous hypothesis about the role of the large-scale flows. The situation is less clear in the puff case in Fig.~\ref{fig:2d01_lsf}. Unlike for higher $\eta$ no robust oblique large-scale flow can be found, neither on arbitrary-$r$ cylinders nor for the wall-integrated field $(\overline{U_x},\overline{U_\theta})$.\\

\red{Let us focus also on the local angle $\alpha$ of the large-scale flow defined by Eq.~(\ref{def_alpha}), in addition to the azimuthal velocity component $\overline{U_\theta}$ alone. Both quantities are functions of $(x, \theta)$ position and time.}
Statistics of $\overline{U_\theta}$ and $\alpha$ have been gathered for all parameters over the numerical grid $(x_i,z_j)$ and over different times. Since we are mainly interested in the values of $\overline{U_\theta}$ at the laminar-turbulent interfaces, we exclude from the statistics fully laminar portions of the flow which would overestimate the statistical weight of the $\overline{U_\theta} \approx 0$ contribution. This is achieved by conditioning all statistics by the additional constraint $|u'_r|/u_{\tau}>0.2$ (which is never fulfilled in laminar zones where $u'_r \approx 0$).
Probability distribution functions (PDFs) for $\left| \overline{U_\theta} \right|$ and $\alpha$ are obtained by considering bins of width $\Delta \overline{U_\theta} = 0.025$ and $\Delta \alpha$=0.25.\\

\begin{figure}[t]
\begin{center}
(a)\includegraphics[width=110mm]{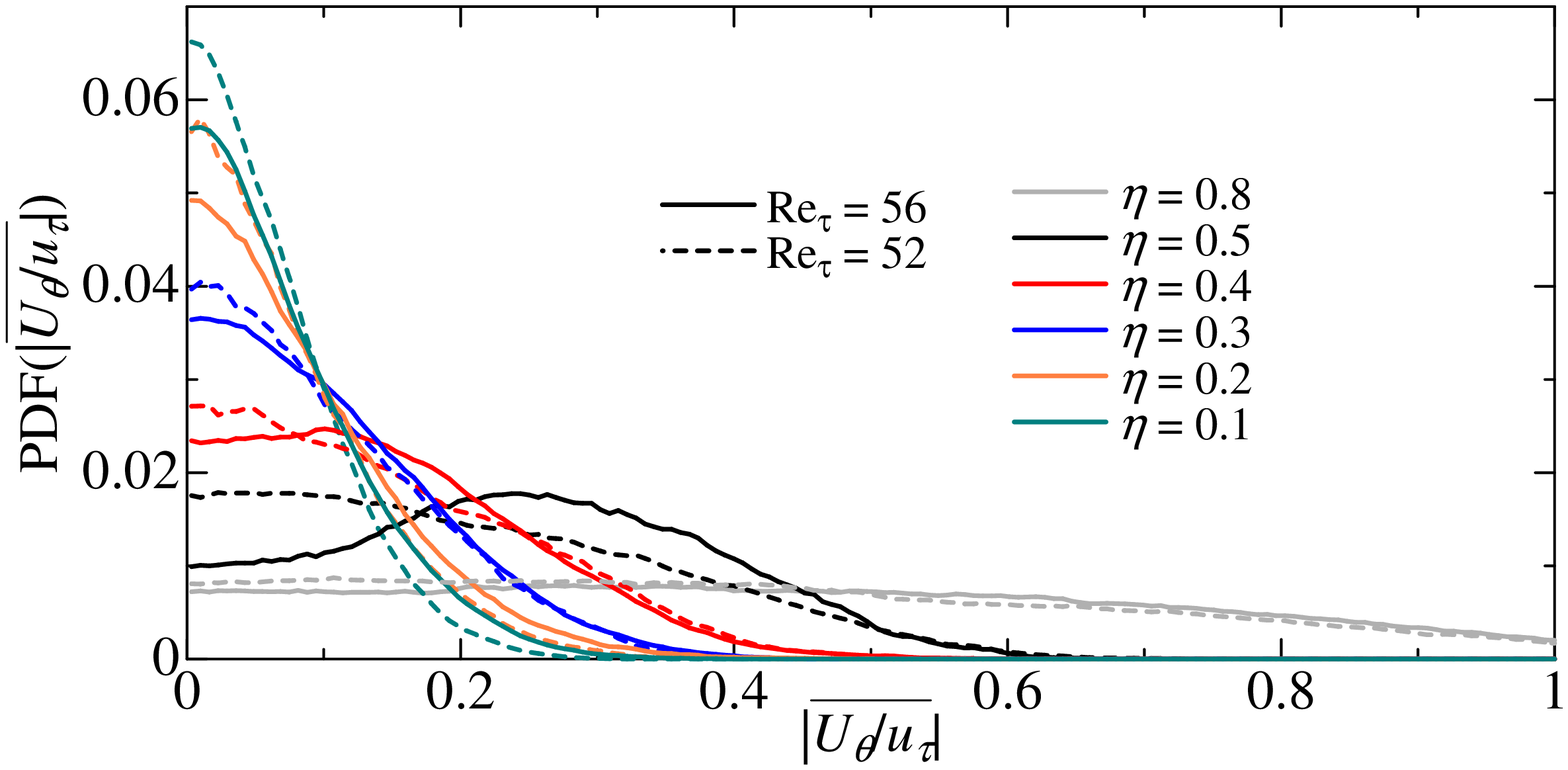}\\
(b)\includegraphics[width=110mm]{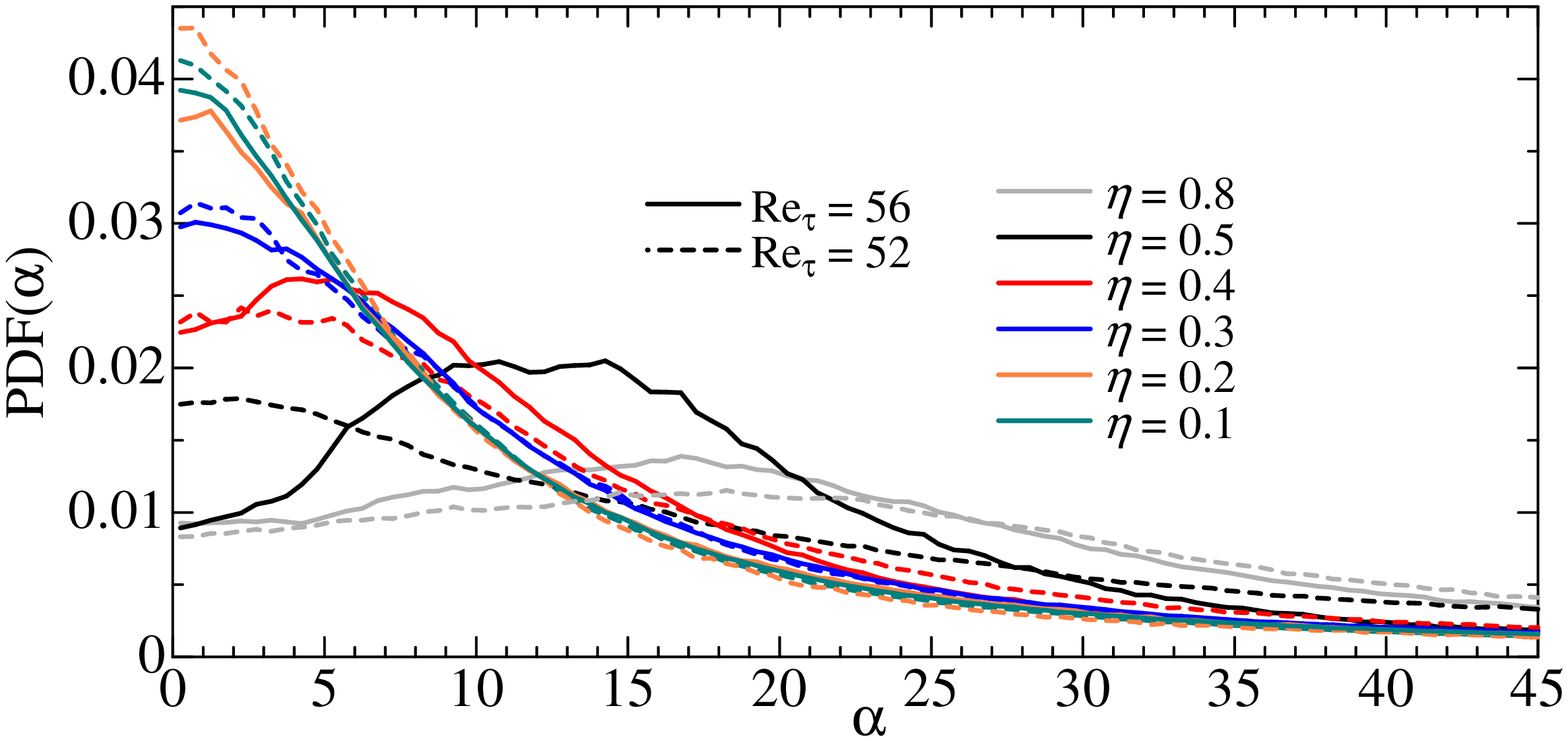}
\vspace{-0.5em}
\end{center}
\caption{PDFs of quantities that characterise the azimuthal large-scale flow and laminar-turbulent interface : (a) $\left| \overline{U_\theta} \right|/u_{\tau}$ and (b) $\alpha$, for ${\rm Re}_{\tau}=52$ and 56, $\eta = 0.1$--0.8.}
\label{fig:pdf}
\end{figure}

We first describe the PDFs of $\left| \overline{U_\theta} \right|$ (normalised by $u_{\tau}$) obtained for several values of $\eta$ and parametrised by ${\rm Re}_{\tau}$, and shown in Fig.~\ref{fig:pdf}(a). For $\eta$=0.1, 0.2 and 0.3, the PDF looks reasonably Gaussian. 
This excludes statistically significant non-zero values of the azimuthal component. Note that we are here only considering the radially-integrated 
azimuthal large-scale velocity component, which in principle does not exclude local weak 'zonal' flows \cite{goldenfeld2016turbulence} with an almost vanishing radial average. For increasing values of $\eta \ge 0.3$, the tendency for the PDF to flatten away from zero becomes stronger, thereby enlarging the range of values of $|\overline{U_\theta}|$. For $\eta=0.4$, a new peak emerges in the PDF from the former tail in the range $\overline{U_\theta}=0.2$--0.3$u_{\tau}$, and this peak overweighs clearly the $\overline{U_\theta}=0$ contribution at $\eta \ge 0.5$. This shows how increasing $\eta$ beyond 0.3 leads with increasing probability to the emergence of a transverse large-scale flow component, interpreted as responsible for the oblique interfaces observed. Focusing on the lowest values of ${\rm Re}_{\tau}$ = 52 and 56 where intermittency has been observed, we next analyse more quantitatively the PDFs of both $|\overline{U_\theta}|/u_{\tau}$ and $\alpha$, parametrised by $\eta$ in the range 0.1--0.8, cf. Figs.~\ref{fig:pdf}(a, b), with the aim of extracting a comprehensive bifurcation diagram as a function of $\eta$. 

\begin{figure}[t]
\begin{center}
(a)\includegraphics[width=100mm]{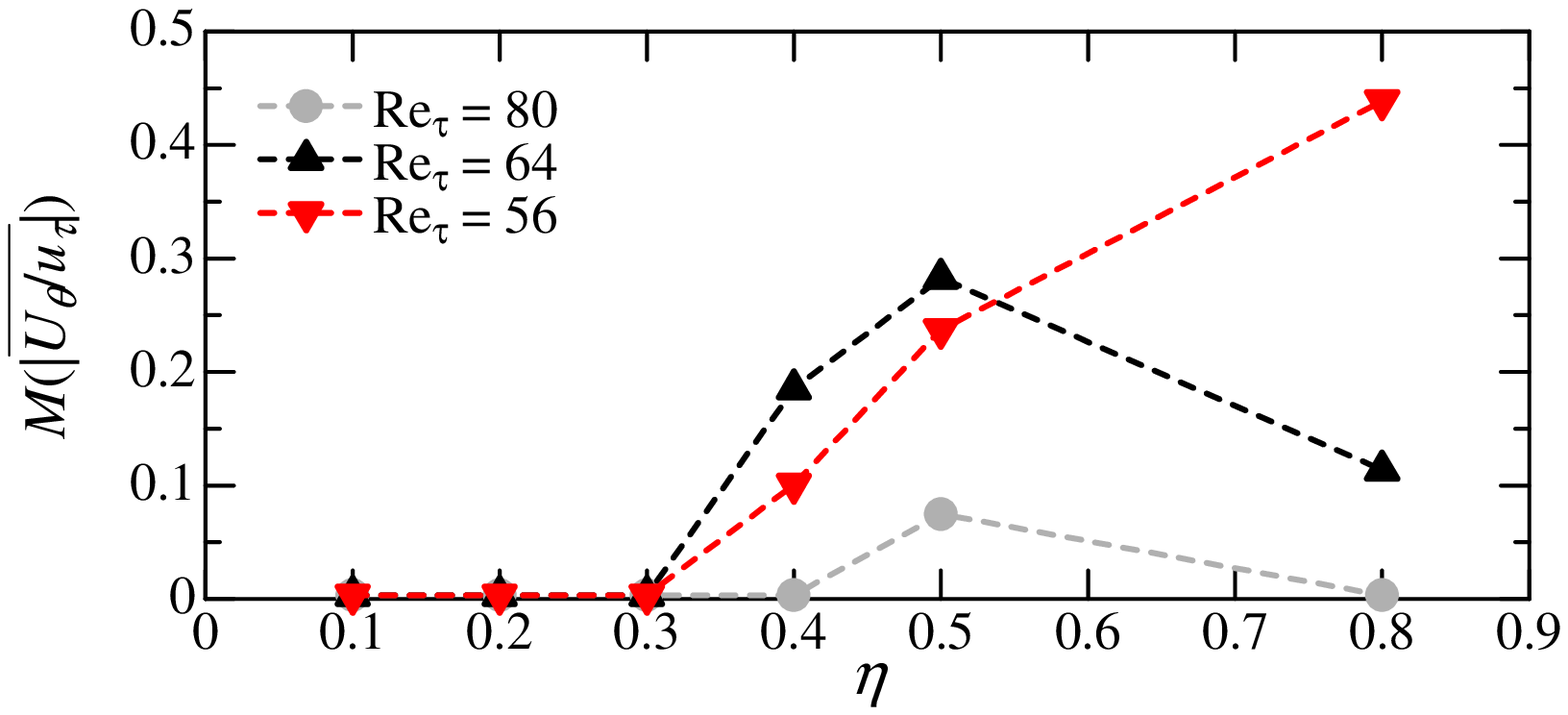}\\
(b)\includegraphics[width=100mm]{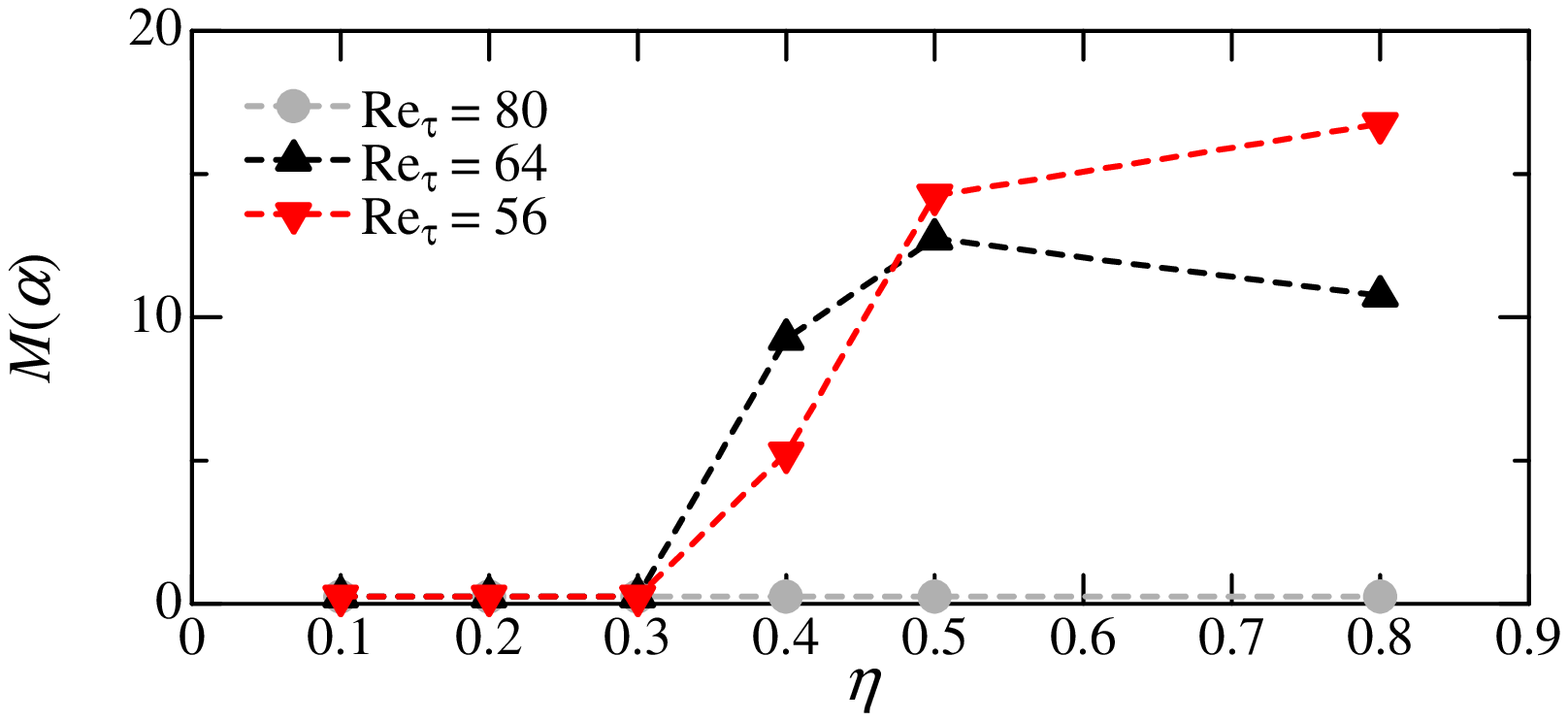}
\vspace{-0.5em}
\end{center}
\caption{Statistical mode extracted from the PDFs in Fig. \ref{fig:pdf} of : (a) $\left| \overline{U_\theta} \right|/u_{\tau}$ and (b) $\alpha$, for ${\rm Re}_{\tau}=52$ and 56, $\eta = 0.1$--0.8. \red{The data suggests a critical value of $\eta=\eta_c$ between 0.3 and 0.4}.}
\label{fig:mode}

\vspace{1.0em}
\begin{center}
~~~\includegraphics[width=100mm]{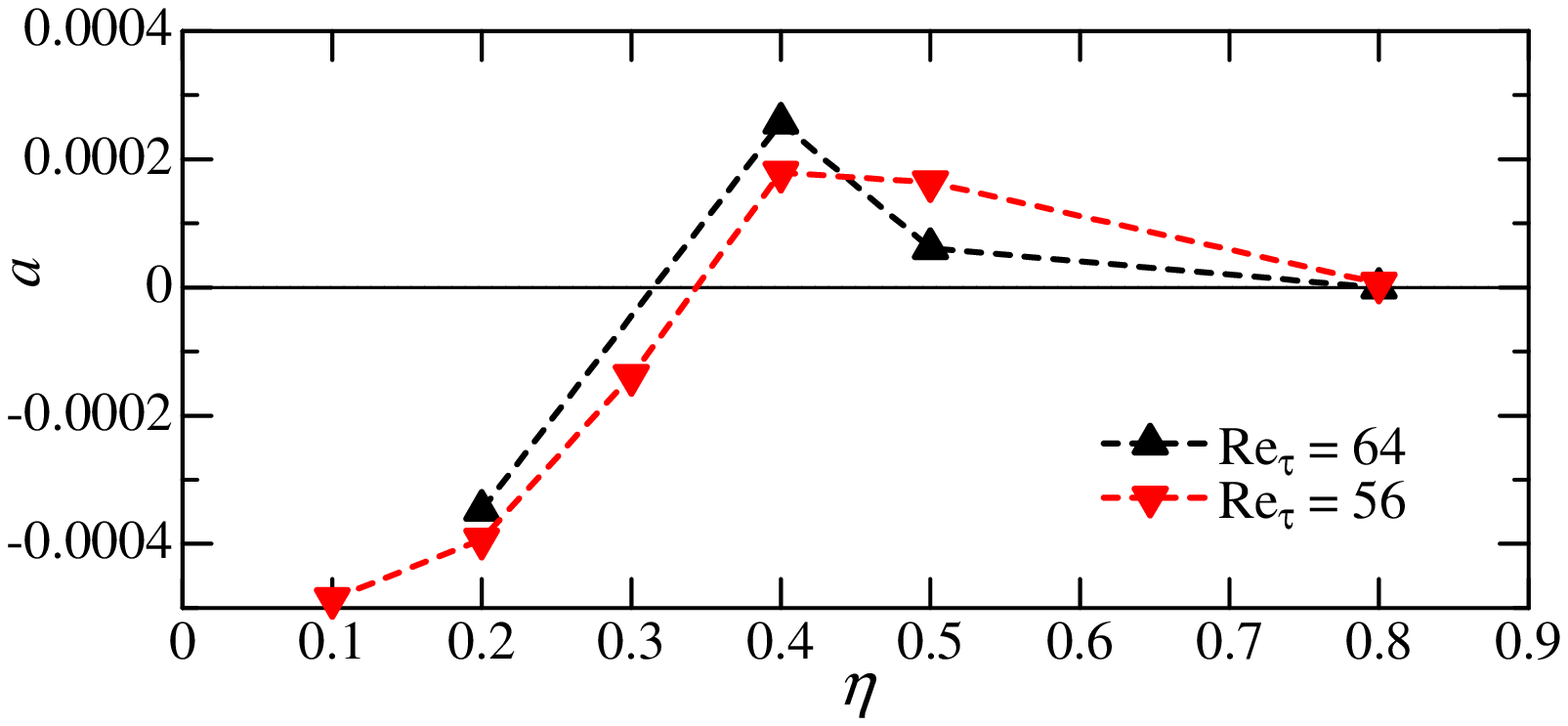}
\vspace{-0.5em}
\end{center}
\caption{Coefficient $a$ in Eq.~(\ref{eq:aa}) extracted from the PDF of the angle $\alpha$.}
\label{fig:aa}
\end{figure}

The transition from a unimodal distribution of both (unsigned) $\overline{U_{\theta}}$ and $\alpha$ \red{cannot be well analysed} using standard statistical moments such as the mean and the variance. Instead we can extract, for both distributions, the statistical mode $M$, i.e. the global maximum of each PDF. They are reported in Fig.~\ref{fig:mode}(a) and \ref{fig:mode}(b), respectively. In both cases the maximum is at 0 for $\eta \le$ 0.3 and jumps to a non-zero value for the next investigated value, i.e. for $\eta \ge$ 0.4. The mode initially increases as $\eta$ increases, however we note a decrease of both $M(\overline{U_{\theta}}/u_{\tau})$ and $M(\alpha)$ for the largest value of $\eta=0.8$. This is due to the flattening of the PDF observed in Fig.~\ref{fig:pdf} due to a wider range of possible angles (not shown). The statistical mode analysis allows to read directly the typical angles of the oblique stripes of Fig.~\ref{fig:mode}(b). However, it does not yield an accurate value of the critical value $\eta_c($Re$_{\tau})$ of $\eta$ at which stripes emerge : all that can be deduced is that $\eta_c$ lies in the range $[0.3:0.4]$. An alternative, specific to the case of unimodal-bimodal transitions, \red{is to measure} the convexity of the PDF near the origin. Suppose that the PDFs in Fig.~\ref{fig:pdf} are even functions $p(x)$ and can be fitted near the origin $x=0$ as 
\begin{eqnarray}
p(x,\eta, {\rm Re}_{\tau})=ax^2+c,
\label{eq:aa}
\end{eqnarray}
with $a=a(\eta, {\rm Re}_{\tau})$ and $c=p(0, \eta$, Re$_{\tau})$. The coefficient $a(\eta, {\rm Re}_{\tau})$ is negative for a unimodal distribution centred at $x=0$ and positive for a bimodal distribution whose maximum lies away from $x=0$. As a consequence, $a(\eta$, Re$_{\tau})=0$ can be used to define $\eta_c(Re_{\tau})$. The values of $a(\eta,Re_{\tau})$ are plotted in Fig. \ref{fig:aa} as functions of $\eta$ parametrised by Re$_{\tau}$ after the PDF of $\alpha$ has been approximated by a quadratic function using a least squares algorithm. Consistently with the mode analysis, $\eta_c$ lies in the interval  $[0.3:0.4]$ for both Re$_{\tau}=$ 64 and 56 (the value Re$_{\tau}$ = 80 has \red{been omitted} as it is too high to sustain oblique stripes). The present data suggests even a slight decrease of $\eta_c$ with Re$_{\tau}$, \red{but} this remains to be verified for a larger number of values of Re$_{\tau}$.

\section{5. Discussion}

The present results have confirmed that annular pipe flow (aPf), parametrised by both its radius ratio $\eta$ and the friction Reynolds number ${\rm Re}_{\tau}$, is a relevant candidate to track the transition from puffs to oblique stripes. The main physical mechanism responsible for this transition is the relaxation of the azimuthal confinement (measured in units of the gap $d$) as $\eta$ increases, allowing for more freedom in the orientation of the large-scale flow occurring at the laminar-turbulent interfaces, should such interfaces exist. Beyond a critical azimuthal extent, a neater selection of orientations occurs, ruled by the mass conservation at the interfaces. As $\eta$ increases from 0 to 1, the following turbulent regimes are encountered : 
\begin{itemize}
\item spatio-temporal intermittency as in circular pipe flow for 0$<\eta \lesssim$ 0.2
\item mixed distribution of straight and helical puffs for 0.2 $\lesssim \eta \lesssim$ 0.4
\item regular patterns of oblique stripes, so-called helical turbulence, for $\eta \gtrsim 0.5$
\item disordered patterns of oblique stripes for $0.5 \ll \eta \lesssim 1$.
\end{itemize}
The last item has not been verified here but is expected to match all observations in extended planar shear flows (with additional effects due to the small wall curvature). This apparent disorganisation occurs when the transverse extent is sufficiently larger than the correlation length of the intermittent regime, at least sufficiently above the critical point Re$^c_{\tau}(\eta)$. The parameter space may also contain new regimes so far unexplored. A statistical analysis has been carried out by focusing entirely on the local structure and orientation of the large-scale flow. Note that other approaches are possible to handle bifurcations from one turbulent regime to another one. When the two regimes of interest are characterised by \red{symmetry breaking} (as is the case here where each oblique stripe violates the $\overline{U_\theta}=0$ symmetry), other order parameters can be considered (see e.g. \cite{Cortet2010}). In the spirit of pattern formation, the statistics of some well-chosen spectral coefficient characteristics of the structures under study can be helpful. The transition from full turbulence to stripe patterns or to puffs has been considered precisely in this manner for various shear flows \cite{TuckermanBarkleyDauchot2008,Moxey10,Tuckerman2014}. Orientation reversals of turbulent stripes in pCf were also monitored by considering the competition between two or more Fourier coefficients corresponding to different wavevectors \cite{RollandManneville2011}.  Modal approaches however show the strong disadvantage of being domain-dependent. We believe that the current statistical approach, more local than modal, is more adapted to  the spatio-temporally intermittent flow regimes encountered here, especially for low $\eta$.\\

While an experimental verification of such flow regimes is called for, it is instructive to discuss whether other shear flow geometries lend themselves easily to bifurcations between intermittently turbulent regimes. The transition from puffs to stripes corresponds essentially, as previously shown, to a transition from one-dimensional to two-dimensional coexistence, via the occurrence of quasi-one-dimensional stripe patterns. Other homotopies can be suggested by similarly tuning the confinement in the direction transverse to the mean flow.  Imposing confinement by sidewalls in an otherwise Cartesian geometry is a possible option. A rectangular duct flow driven by a pressure gradient, with a cross-section of dimensions $L_y \times L_z$, can be parametrised both by a Reynolds number and an aspect ratio $A=L_z/L_y$ ($A \ge$ 1 by convention). The geometry of $A=1$ corresponds to a square duct whereas $A \rightarrow \infty$ is equivalent to plane Poiseuille flow. The laminar profile is known to be linearly stable for all values of Re$_\tau$ of interest here. This parametric problem has been considered numerically in Ref.~\cite{TakeishiJFM2015}. Again straight puffs have been identified for $A \le 3$ whereas spots with somewhat oblique interfaces have been visualised for $A \ge 4$. The authors \red{have reported} a peculiarity of the confinement by \red{solid sidewalls} : permanent local relaminarisation at the \red{sidewalls} affecting the localisation of the turbulent structure. This is consistent with the recent experimental observations of spots in a duct flow with aspect ratio $A=7.5$ \cite{LemoultEPJE2014}. While this system has the advantage of dealing with flat walls only, the local relaminarisation at the sidewalls is interpreted as an additional complication obscuring the transition from puffs to oblique stripes. \\

 Other examples closer to the present case of aPf share a common geometry but differ in the way energy is injected into the flow. The Taylor-Couette system, where the two coaxial cylinders rotate with different frequencies $\Omega_i$ and $\Omega_o$ in the absence of \red{an axial} pressure gradient, is such an example. It is known that for $\eta \rightarrow 1$, the case of exact counter-rotation $\mu=\Omega_o/\Omega_i=-1$ corresponds to plane Couette flow, which has a linearly stable base flow. It has been experimentally verified for $\eta$ slightly below 1 that  low-$Re$ regimes with $\mu \approx -1$ feature two-dimensional intermittent arrangements of oblique stripes with both positive and negative angles \cite{Prigent02,BorreroPRE2010,AvilaHof2013}. Reducing $\eta$ with fixed $\mu$, however, leads to changes in the stability of the base flow. \red{Other regimes} such as those featuring interpenetrating spirals \cite{Andereck1986} enter the bifurcation diagram and make the continuation from stripes to puffs unlikely. It is an open question whether other paths in an enlarged parameter space could lead to such a transition.\\

Finally, an interesting candidate for the continuation from stripes to puffs is the sliding Couette flow (sCf), where the outer cylinder is fixed and the inner one moves axially with a constant velocity (see e.g. \cite{DeguchiNagata2011}). This flow is thought to be equivalent to pCf in the $\eta \rightarrow 1$ limit and again to pipe flow in the vanishing $\eta$ limit. It has been verified numerically \cite{KuniiICTAM2016} that this system bridges puffs to spots in a way apparently similar to the present system. Whether and how the homotopies in sCf and aPf really differ remains an open question. However, aPf shows the important advantage of being easier to achieve experimentally as it does not include any motion of the solid walls, only a pressure gradient imposed on a fixed geometry.

\vspace{0.5cm} \noindent {\bf Acknowledgements}\\

\begin{acknowledgments}
T.I. was supported by a Grant-in-Aid from JSPS (Japan Society for the Promotion of Science) Fellowship \#26-7477. 
This work was partially supported by JSPS KAKENHI Grant Number 16H06066 and 16H00813.
The present simulations were performed on supercomputers at the Cyberscience Centre of Tohoku University and at the Cybermedia Centre of Osaka University. 
We thank JSPS, CNRS (Centre National de la Recherche Scientifique), and RIMS (Research Institute for \red{Mathematical Sciences}) for additional \red{travel support}. 
\end{acknowledgments}

\bibliography{ishidaAPF2}

\begin{thebibliography}{32}%
\makeatletter
\providecommand \@ifxundefined [1]{%
 \@ifx{#1\undefined}
}%
\providecommand \@ifnum [1]{%
 \ifnum #1\expandafter \@firstoftwo
 \else \expandafter \@secondoftwo
 \fi
}%
\providecommand \@ifx [1]{%
 \ifx #1\expandafter \@firstoftwo
 \else \expandafter \@secondoftwo
 \fi
}%
\providecommand \natexlab [1]{#1}%
\providecommand \enquote  [1]{``#1''}%
\providecommand \bibnamefont  [1]{#1}%
\providecommand \bibfnamefont [1]{#1}%
\providecommand \citenamefont [1]{#1}%
\providecommand \href@noop [0]{\@secondoftwo}%
\providecommand \href [0]{\begingroup \@sanitize@url \@href}%
\providecommand \@href[1]{\@@startlink{#1}\@@href}%
\providecommand \@@href[1]{\endgroup#1\@@endlink}%
\providecommand \@sanitize@url [0]{\catcode `\\12\catcode `\$12\catcode
  `\&12\catcode `\#12\catcode `\^12\catcode `\_12\catcode `\%12\relax}%
\providecommand \@@startlink[1]{}%
\providecommand \@@endlink[0]{}%
\providecommand \url  [0]{\begingroup\@sanitize@url \@url }%
\providecommand \@url [1]{\endgroup\@href {#1}{\urlprefix }}%
\providecommand \urlprefix  [0]{URL }%
\providecommand \Eprint [0]{\href }%
\providecommand \doibase [0]{http://dx.doi.org/}%
\providecommand \selectlanguage [0]{\@gobble}%
\providecommand \bibinfo  [0]{\@secondoftwo}%
\providecommand \bibfield  [0]{\@secondoftwo}%
\providecommand \translation [1]{[#1]}%
\providecommand \BibitemOpen [0]{}%
\providecommand \bibitemStop [0]{}%
\providecommand \bibitemNoStop [0]{.\EOS\space}%
\providecommand \EOS [0]{\spacefactor3000\relax}%
\providecommand \BibitemShut  [1]{\csname bibitem#1\endcsname}%
\let\auto@bib@innerbib\@empty
\bibitem [{\citenamefont {Manneville}(2016)}]{Manneville2016}%
  \BibitemOpen
  \bibfield  {author} {\bibinfo {author} {\bibfnamefont {Paul}\ \bibnamefont
  {Manneville}},\ }\bibfield  {title} {\enquote {\bibinfo {title} {Transition
  to turbulence in wall-bounded flows: Where do we stand?}}\ }\href@noop {}
  {\bibfield  {journal} {\bibinfo  {journal} {Mechanical Engineering Reviews}\
  } (\bibinfo {year} {2016})}\BibitemShut {NoStop}%
\bibitem [{\citenamefont {Wygnanski}\ and\ \citenamefont
  {Champagne}(1973)}]{Wygnanski73}%
  \BibitemOpen
  \bibfield  {author} {\bibinfo {author} {\bibfnamefont {I.~J.}\ \bibnamefont
  {Wygnanski}}\ and\ \bibinfo {author} {\bibfnamefont {F.~H.}\ \bibnamefont
  {Champagne}},\ }\bibfield  {title} {\enquote {\bibinfo {title} {On transition
  in a pipe. {P}art 1. {T}he origin of puffs and slugs and the flow in a
  turbulent slug},}\ }\href@noop {} {\bibfield  {journal} {\bibinfo  {journal}
  {J. Fluid Mech.}\ }\textbf {\bibinfo {volume} {59}},\ \bibinfo {pages}
  {281--335} (\bibinfo {year} {1973})}\BibitemShut {NoStop}%
\bibitem [{\citenamefont {Avila}\ \emph {et~al.}(2011)\citenamefont {Avila},
  \citenamefont {Moxey}, \citenamefont {de~Lozar}, \citenamefont {Avila},
  \citenamefont {Barkley},\ and\ \citenamefont {Hof}}]{Avila2011}%
  \BibitemOpen
  \bibfield  {author} {\bibinfo {author} {\bibfnamefont {K.}~\bibnamefont
  {Avila}}, \bibinfo {author} {\bibfnamefont {D.}~\bibnamefont {Moxey}},
  \bibinfo {author} {\bibfnamefont {A.}~\bibnamefont {de~Lozar}}, \bibinfo
  {author} {\bibfnamefont {M.}~\bibnamefont {Avila}}, \bibinfo {author}
  {\bibfnamefont {D.}~\bibnamefont {Barkley}}, \ and\ \bibinfo {author}
  {\bibfnamefont {B.}~\bibnamefont {Hof}},\ }\bibfield  {title} {\enquote
  {\bibinfo {title} {The onset of turbulence in pipe flow},}\ }\href@noop {}
  {\bibfield  {journal} {\bibinfo  {journal} {Science}\ }\textbf {\bibinfo
  {volume} {333}},\ \bibinfo {pages} {192--196} (\bibinfo {year}
  {2011})}\BibitemShut {NoStop}%
\bibitem [{\citenamefont {Barkley}(2016)}]{Barkley2016}%
  \BibitemOpen
  \bibfield  {author} {\bibinfo {author} {\bibfnamefont {Dwight}\ \bibnamefont
  {Barkley}},\ }\bibfield  {title} {\enquote {\bibinfo {title} {Theoretical
  perspective on the route to turbulence in a pipe},}\ }\href@noop {}
  {\bibfield  {journal} {\bibinfo  {journal} {Journal of Fluid Mechanics}\
  }\textbf {\bibinfo {volume} {803}} (\bibinfo {year} {2016})}\BibitemShut
  {NoStop}%
\bibitem [{\citenamefont {Wesfreid}\ and\ \citenamefont
  {Klotz}(2016)}]{Klotz2016}%
  \BibitemOpen
  \bibfield  {author} {\bibinfo {author} {\bibfnamefont {Jose~Eduardo}\
  \bibnamefont {Wesfreid}}\ and\ \bibinfo {author} {\bibfnamefont {Lukasz}\
  \bibnamefont {Klotz}},\ }\bibfield  {title} {\enquote {\bibinfo {title}
  {Subcritical transition to turbulence in {C}ouette-{P}oiseuille flow},}\
  }\href@noop {} {\bibfield  {journal} {\bibinfo  {journal} {Bulletin of the
  American Physical Society}\ }\textbf {\bibinfo {volume} {61}} (\bibinfo
  {year} {2016})}\BibitemShut {NoStop}%
\bibitem [{\citenamefont {Prigent}\ \emph {et~al.}(2002)\citenamefont
  {Prigent}, \citenamefont {Gr{\'e}goire}, \citenamefont {Chat{\'e}},
  \citenamefont {Dauchot},\ and\ \citenamefont {van Saarloos}}]{Prigent02}%
  \BibitemOpen
  \bibfield  {author} {\bibinfo {author} {\bibfnamefont {A.}~\bibnamefont
  {Prigent}}, \bibinfo {author} {\bibfnamefont {G.}~\bibnamefont
  {Gr{\'e}goire}}, \bibinfo {author} {\bibfnamefont {H.}~\bibnamefont
  {Chat{\'e}}}, \bibinfo {author} {\bibfnamefont {O.}~\bibnamefont {Dauchot}},
  \ and\ \bibinfo {author} {\bibfnamefont {W.}~\bibnamefont {van Saarloos}},\
  }\bibfield  {title} {\enquote {\bibinfo {title} {Large-scale
  finite-wavelength modulation within turbulent shear flows},}\ }\href@noop {}
  {\bibfield  {journal} {\bibinfo  {journal} {Phys. Rev. Lett.}\ }\textbf
  {\bibinfo {volume} {89}},\ \bibinfo {pages} {014501} (\bibinfo {year}
  {2002})}\BibitemShut {NoStop}%
\bibitem [{\citenamefont {Tsukahara}\ \emph {et~al.}(2005)\citenamefont
  {Tsukahara}, \citenamefont {Seki}, \citenamefont {Kawamura},\ and\
  \citenamefont {Tochio}}]{Tsukahara2005}%
  \BibitemOpen
  \bibfield  {author} {\bibinfo {author} {\bibfnamefont {T.}~\bibnamefont
  {Tsukahara}}, \bibinfo {author} {\bibfnamefont {Y.}~\bibnamefont {Seki}},
  \bibinfo {author} {\bibfnamefont {H.}~\bibnamefont {Kawamura}}, \ and\
  \bibinfo {author} {\bibfnamefont {D.}~\bibnamefont {Tochio}},\ }\bibfield
  {title} {\enquote {\bibinfo {title} {{DNS} of turbulent channel flow at very
  low {R}eynolds numbers},}\ }in\ \href@noop {} {\emph {\bibinfo {booktitle}
  {Proc. Fourth Int. Symp. on Turbulence and Shear Flow Phenomena}}},\ \bibinfo
  {editor} {edited by\ \bibinfo {editor} {\bibfnamefont {J.~A. C.~{\em
  et~al.\/}}\ \bibnamefont {Humphrey}}}\ (\bibinfo {address} {Williamsburg,
  USA},\ \bibinfo {year} {2005})\ pp.\ \bibinfo {pages} {935--940},\ \Eprint
  {http://arxiv.org/abs/arXiv preprint: 1406.0248} {arXiv preprint: 1406.0248}
  \BibitemShut {NoStop}%
\bibitem [{\citenamefont {Duguet}\ \emph {et~al.}(2010)\citenamefont {Duguet},
  \citenamefont {Schlatter},\ and\ \citenamefont {Henningson}}]{Duguet2010}%
  \BibitemOpen
  \bibfield  {author} {\bibinfo {author} {\bibfnamefont {Y.}~\bibnamefont
  {Duguet}}, \bibinfo {author} {\bibfnamefont {P.}~\bibnamefont {Schlatter}}, \
  and\ \bibinfo {author} {\bibfnamefont {D.~S.}\ \bibnamefont {Henningson}},\
  }\bibfield  {title} {\enquote {\bibinfo {title} {Formation of turbulent
  patterns near the onset of transition in plane {C}ouette flow},}\ }\href@noop
  {} {\bibfield  {journal} {\bibinfo  {journal} {J. Fluid Mech.}\ }\textbf
  {\bibinfo {volume} {650}},\ \bibinfo {pages} {119--129} (\bibinfo {year}
  {2010})}\BibitemShut {NoStop}%
\bibitem [{\citenamefont {Lemoult}\ \emph {et~al.}(2016)\citenamefont
  {Lemoult}, \citenamefont {Shi}, \citenamefont {Avila}, \citenamefont
  {Jalikop}, \citenamefont {Avila},\ and\ \citenamefont {Hof}}]{Lemoult2016}%
  \BibitemOpen
  \bibfield  {author} {\bibinfo {author} {\bibfnamefont {Gr{\'e}goire}\
  \bibnamefont {Lemoult}}, \bibinfo {author} {\bibfnamefont {Liang}\
  \bibnamefont {Shi}}, \bibinfo {author} {\bibfnamefont {Kerstin}\ \bibnamefont
  {Avila}}, \bibinfo {author} {\bibfnamefont {Shreyas~V}\ \bibnamefont
  {Jalikop}}, \bibinfo {author} {\bibfnamefont {Marc}\ \bibnamefont {Avila}}, \
  and\ \bibinfo {author} {\bibfnamefont {Bj{\"o}rn}\ \bibnamefont {Hof}},\
  }\bibfield  {title} {\enquote {\bibinfo {title} {Directed percolation phase
  transition to sustained turbulence in {C}ouette flow},}\ }\href@noop {}
  {\bibfield  {journal} {\bibinfo  {journal} {Nature Physics}\ } (\bibinfo
  {year} {2016})}\BibitemShut {NoStop}%
\bibitem [{\citenamefont {Barkley}(2011)}]{Barkley11}%
  \BibitemOpen
  \bibfield  {author} {\bibinfo {author} {\bibfnamefont {D.}~\bibnamefont
  {Barkley}},\ }\bibfield  {title} {\enquote {\bibinfo {title} {Simplifying the
  complexity of pipe flow},}\ }\href@noop {} {\bibfield  {journal} {\bibinfo
  {journal} {Phys. Rev. E}\ }\textbf {\bibinfo {volume} {84}},\ \bibinfo
  {pages} {016309} (\bibinfo {year} {2011})}\BibitemShut {NoStop}%
\bibitem [{\citenamefont {Walker}\ \emph {et~al.}(1957)\citenamefont {Walker},
  \citenamefont {Whan},\ and\ \citenamefont {Rothfus}}]{Walker57}%
  \BibitemOpen
  \bibfield  {author} {\bibinfo {author} {\bibfnamefont {J.~E.}\ \bibnamefont
  {Walker}}, \bibinfo {author} {\bibfnamefont {G.~A.}\ \bibnamefont {Whan}}, \
  and\ \bibinfo {author} {\bibfnamefont {R.~R.}\ \bibnamefont {Rothfus}},\
  }\bibfield  {title} {\enquote {\bibinfo {title} {Fluid friction in
  noncircular ducts},}\ }\href@noop {} {\bibfield  {journal} {\bibinfo
  {journal} {AIChE Journal}\ }\textbf {\bibinfo {volume} {3}},\ \bibinfo
  {pages} {484--489} (\bibinfo {year} {1957})}\BibitemShut {NoStop}%
\bibitem [{\citenamefont {Heaton}(2008)}]{Heaton08}%
  \BibitemOpen
  \bibfield  {author} {\bibinfo {author} {\bibfnamefont {C.~J.}\ \bibnamefont
  {Heaton}},\ }\bibfield  {title} {\enquote {\bibinfo {title} {Linear
  instability of annular {P}oiseuille flow},}\ }\href@noop {} {\bibfield
  {journal} {\bibinfo  {journal} {J. Fluid Mech.}\ }\textbf {\bibinfo {volume}
  {610}},\ \bibinfo {pages} {391--406} (\bibinfo {year} {2008})}\BibitemShut
  {NoStop}%
\bibitem [{\citenamefont {Ishida}\ \emph {et~al.}(2016)\citenamefont {Ishida},
  \citenamefont {Duguet},\ and\ \citenamefont {Tsukahara}}]{IDT2016}%
  \BibitemOpen
  \bibfield  {author} {\bibinfo {author} {\bibfnamefont {Takahiro}\
  \bibnamefont {Ishida}}, \bibinfo {author} {\bibfnamefont {Yohann}\
  \bibnamefont {Duguet}}, \ and\ \bibinfo {author} {\bibfnamefont {Takahiro}\
  \bibnamefont {Tsukahara}},\ }\bibfield  {title} {\enquote {\bibinfo {title}
  {Transitional structures in annular {P}oiseuille flow depending on radius
  ratio},}\ }\href@noop {} {\bibfield  {journal} {\bibinfo  {journal} {Journal
  of Fluid Mechanics}\ }\textbf {\bibinfo {volume} {794}},\ \bibinfo {pages}
  {R2} (\bibinfo {year} {2016})}\BibitemShut {NoStop}%
\bibitem [{\citenamefont {Abe}\ \emph {et~al.}(2001)\citenamefont {Abe},
  \citenamefont {Kawamura},\ and\ \citenamefont {Matsuo}}]{Abe01}%
  \BibitemOpen
  \bibfield  {author} {\bibinfo {author} {\bibfnamefont {H.}~\bibnamefont
  {Abe}}, \bibinfo {author} {\bibfnamefont {H.}~\bibnamefont {Kawamura}}, \
  and\ \bibinfo {author} {\bibfnamefont {Y.}~\bibnamefont {Matsuo}},\
  }\bibfield  {title} {\enquote {\bibinfo {title} {Direct numerical simulation
  of a fully developed turbulent channel flow with respect to the {R}eynolds
  number dependence},}\ }\href@noop {} {\bibfield  {journal} {\bibinfo
  {journal} {J. Fluids Eng.}\ }\textbf {\bibinfo {volume} {123}},\ \bibinfo
  {pages} {382--393} (\bibinfo {year} {2001})}\BibitemShut {NoStop}%
\bibitem [{\citenamefont {Philip}\ and\ \citenamefont
  {Manneville}(2011)}]{PhilipManneville2010}%
  \BibitemOpen
  \bibfield  {author} {\bibinfo {author} {\bibfnamefont {Jimmy}\ \bibnamefont
  {Philip}}\ and\ \bibinfo {author} {\bibfnamefont {Paul}\ \bibnamefont
  {Manneville}},\ }\bibfield  {title} {\enquote {\bibinfo {title} {From
  temporal to spatiotemporal dynamics in transitional plane {C}ouette flow},}\
  }\href@noop {} {\bibfield  {journal} {\bibinfo  {journal} {Physical Review
  E}\ }\textbf {\bibinfo {volume} {83}},\ \bibinfo {pages} {036308} (\bibinfo
  {year} {2011})}\BibitemShut {NoStop}%
\bibitem [{\citenamefont {Moxey}\ and\ \citenamefont
  {Barkley}(2010)}]{Moxey10}%
  \BibitemOpen
  \bibfield  {author} {\bibinfo {author} {\bibfnamefont {D.}~\bibnamefont
  {Moxey}}\ and\ \bibinfo {author} {\bibfnamefont {D.}~\bibnamefont
  {Barkley}},\ }\bibfield  {title} {\enquote {\bibinfo {title} {Distinct
  large-scale turbulent-laminar states in transitional pipe flow},}\
  }\href@noop {} {\bibfield  {journal} {\bibinfo  {journal} {PNAS}\ }\textbf
  {\bibinfo {volume} {107}},\ \bibinfo {pages} {8091--8096} (\bibinfo {year}
  {2010})}\BibitemShut {NoStop}%
\bibitem [{\citenamefont {Shimizu}\ \emph {et~al.}(2014)\citenamefont
  {Shimizu}, \citenamefont {Manneville}, \citenamefont {Duguet},\ and\
  \citenamefont {Kawahara}}]{Shimizu2014}%
  \BibitemOpen
  \bibfield  {author} {\bibinfo {author} {\bibfnamefont {M.}~\bibnamefont
  {Shimizu}}, \bibinfo {author} {\bibfnamefont {P.}~\bibnamefont {Manneville}},
  \bibinfo {author} {\bibfnamefont {Y.}~\bibnamefont {Duguet}}, \ and\ \bibinfo
  {author} {\bibfnamefont {G.}~\bibnamefont {Kawahara}},\ }\bibfield  {title}
  {\enquote {\bibinfo {title} {Splitting of a turbulent puff in pipe flow},}\
  }\href@noop {} {\bibfield  {journal} {\bibinfo  {journal} {Fluid Dyn. Res.}\
  }\textbf {\bibinfo {volume} {46}},\ \bibinfo {pages} {061403} (\bibinfo
  {year} {2014})}\BibitemShut {NoStop}%
\bibitem [{\citenamefont {Xiong}\ \emph {et~al.}(2015)\citenamefont {Xiong},
  \citenamefont {Tao}, \citenamefont {Chen},\ and\ \citenamefont
  {Brandt}}]{Xiong15}%
  \BibitemOpen
  \bibfield  {author} {\bibinfo {author} {\bibfnamefont {X.}~\bibnamefont
  {Xiong}}, \bibinfo {author} {\bibfnamefont {J.}~\bibnamefont {Tao}}, \bibinfo
  {author} {\bibfnamefont {S.}~\bibnamefont {Chen}}, \ and\ \bibinfo {author}
  {\bibfnamefont {L.}~\bibnamefont {Brandt}},\ }\bibfield  {title} {\enquote
  {\bibinfo {title} {Turbulent bands in plane-{P}oiseuille flow at moderate
  {R}eynolds numbers},}\ }\href@noop {} {\bibfield  {journal} {\bibinfo
  {journal} {Phys. Fluids}\ ,\ \bibinfo {pages} {041702}} (\bibinfo {year}
  {2015})}\BibitemShut {NoStop}%
\bibitem [{\citenamefont {Duguet}\ and\ \citenamefont
  {Schlatter}(2013)}]{DS2013}%
  \BibitemOpen
  \bibfield  {author} {\bibinfo {author} {\bibfnamefont {Y.}~\bibnamefont
  {Duguet}}\ and\ \bibinfo {author} {\bibfnamefont {P.}~\bibnamefont
  {Schlatter}},\ }\bibfield  {title} {\enquote {\bibinfo {title} {Oblique
  laminar-turbulent interfaces in plane shear flows},}\ }\href@noop {}
  {\bibfield  {journal} {\bibinfo  {journal} {Phys. Rev. Lett.}\ }\textbf
  {\bibinfo {volume} {110}},\ \bibinfo {pages} {034502} (\bibinfo {year}
  {2013})}\BibitemShut {NoStop}%
\bibitem [{\citenamefont {Hamilton}\ \emph {et~al.}(1995)\citenamefont
  {Hamilton}, \citenamefont {Kim},\ and\ \citenamefont {Waleffe}}]{HKW1995}%
  \BibitemOpen
  \bibfield  {author} {\bibinfo {author} {\bibfnamefont {James~M}\ \bibnamefont
  {Hamilton}}, \bibinfo {author} {\bibfnamefont {John}\ \bibnamefont {Kim}}, \
  and\ \bibinfo {author} {\bibfnamefont {Fabian}\ \bibnamefont {Waleffe}},\
  }\bibfield  {title} {\enquote {\bibinfo {title} {Regeneration mechanisms of
  near-wall turbulence structures},}\ }\href@noop {} {\bibfield  {journal}
  {\bibinfo  {journal} {Journal of Fluid Mechanics}\ }\textbf {\bibinfo
  {volume} {287}},\ \bibinfo {pages} {317--348} (\bibinfo {year}
  {1995})}\BibitemShut {NoStop}%
\bibitem [{\citenamefont {Goldenfeld}\ and\ \citenamefont
  {Shih}(2016)}]{goldenfeld2016turbulence}%
  \BibitemOpen
  \bibfield  {author} {\bibinfo {author} {\bibfnamefont {Nigel}\ \bibnamefont
  {Goldenfeld}}\ and\ \bibinfo {author} {\bibfnamefont {Hong-Yan}\ \bibnamefont
  {Shih}},\ }\bibfield  {title} {\enquote {\bibinfo {title} {Turbulence as a
  problem in non-equilibrium statistical mechanics},}\ }\href@noop {}
  {\bibfield  {journal} {\bibinfo  {journal} {Journal of Statistical Physics}\
  ,\ \bibinfo {pages} {1--20}} (\bibinfo {year} {2016})}\BibitemShut {NoStop}%
\bibitem [{\citenamefont {Cortet}\ \emph {et~al.}(2010)\citenamefont {Cortet},
  \citenamefont {Chiffaudel}, \citenamefont {Daviaud},\ and\ \citenamefont
  {Dubrulle}}]{Cortet2010}%
  \BibitemOpen
  \bibfield  {author} {\bibinfo {author} {\bibfnamefont {P-P}\ \bibnamefont
  {Cortet}}, \bibinfo {author} {\bibfnamefont {A}~\bibnamefont {Chiffaudel}},
  \bibinfo {author} {\bibfnamefont {F}~\bibnamefont {Daviaud}}, \ and\ \bibinfo
  {author} {\bibfnamefont {B}~\bibnamefont {Dubrulle}},\ }\bibfield  {title}
  {\enquote {\bibinfo {title} {Experimental evidence of a phase transition in a
  closed turbulent flow},}\ }\href@noop {} {\bibfield  {journal} {\bibinfo
  {journal} {Physical Review Letters}\ }\textbf {\bibinfo {volume} {105}},\
  \bibinfo {pages} {214501} (\bibinfo {year} {2010})}\BibitemShut {NoStop}%
\bibitem [{\citenamefont {Tuckerman}\ \emph {et~al.}(2008)\citenamefont
  {Tuckerman}, \citenamefont {Barkley},\ and\ \citenamefont
  {Dauchot}}]{TuckermanBarkleyDauchot2008}%
  \BibitemOpen
  \bibfield  {author} {\bibinfo {author} {\bibfnamefont {Laurette~S}\
  \bibnamefont {Tuckerman}}, \bibinfo {author} {\bibfnamefont {Dwight}\
  \bibnamefont {Barkley}}, \ and\ \bibinfo {author} {\bibfnamefont {Olivier}\
  \bibnamefont {Dauchot}},\ }\bibfield  {title} {\enquote {\bibinfo {title}
  {Statistical analysis of the transition to turbulent-laminar banded patterns
  in plane {C}ouette flow},}\ }\href@noop {} {\bibfield  {journal} {\bibinfo
  {journal} {Journal of Physics: Conference Series}\ }\textbf {\bibinfo
  {volume} {137}},\ \bibinfo {pages} {012029} (\bibinfo {year}
  {2008})}\BibitemShut {NoStop}%
\bibitem [{\citenamefont {Tuckerman}\ \emph {et~al.}(2014)\citenamefont
  {Tuckerman}, \citenamefont {Kreilos}, \citenamefont {Schrobsdorff},
  \citenamefont {Schneider},\ and\ \citenamefont {Gibson}}]{Tuckerman2014}%
  \BibitemOpen
  \bibfield  {author} {\bibinfo {author} {\bibfnamefont {L.~S.}\ \bibnamefont
  {Tuckerman}}, \bibinfo {author} {\bibfnamefont {T.}~\bibnamefont {Kreilos}},
  \bibinfo {author} {\bibfnamefont {H.}~\bibnamefont {Schrobsdorff}}, \bibinfo
  {author} {\bibfnamefont {T.~M.}\ \bibnamefont {Schneider}}, \ and\ \bibinfo
  {author} {\bibfnamefont {J.~F.}\ \bibnamefont {Gibson}},\ }\bibfield  {title}
  {\enquote {\bibinfo {title} {Turbulent-laminar patterns in plane {P}oiseuille
  flow},}\ }\href@noop {} {\bibfield  {journal} {\bibinfo  {journal} {Phys.
  Fluids}\ }\textbf {\bibinfo {volume} {26}},\ \bibinfo {pages} {114103}
  (\bibinfo {year} {2014})}\BibitemShut {NoStop}%
\bibitem [{\citenamefont {Rolland}\ and\ \citenamefont
  {Manneville}(2011)}]{RollandManneville2011}%
  \BibitemOpen
  \bibfield  {author} {\bibinfo {author} {\bibfnamefont {Joran}\ \bibnamefont
  {Rolland}}\ and\ \bibinfo {author} {\bibfnamefont {Paul}\ \bibnamefont
  {Manneville}},\ }\bibfield  {title} {\enquote {\bibinfo {title} {Pattern
  fluctuations in transitional plane {C}ouette flow},}\ }\href@noop {}
  {\bibfield  {journal} {\bibinfo  {journal} {Journal of Statistical Physics}\
  }\textbf {\bibinfo {volume} {142}},\ \bibinfo {pages} {577--591} (\bibinfo
  {year} {2011})}\BibitemShut {NoStop}%
\bibitem [{\citenamefont {Takeishi}\ \emph {et~al.}(2015)\citenamefont
  {Takeishi}, \citenamefont {Kawahara}, \citenamefont {Wakabayashi},
  \citenamefont {Uhlmann},\ and\ \citenamefont {Pinelli}}]{TakeishiJFM2015}%
  \BibitemOpen
  \bibfield  {author} {\bibinfo {author} {\bibfnamefont {Keisuke}\ \bibnamefont
  {Takeishi}}, \bibinfo {author} {\bibfnamefont {Genta}\ \bibnamefont
  {Kawahara}}, \bibinfo {author} {\bibfnamefont {Hiroki}\ \bibnamefont
  {Wakabayashi}}, \bibinfo {author} {\bibfnamefont {Markus}\ \bibnamefont
  {Uhlmann}}, \ and\ \bibinfo {author} {\bibfnamefont {Alfredo}\ \bibnamefont
  {Pinelli}},\ }\bibfield  {title} {\enquote {\bibinfo {title} {Localized
  turbulence structures in transitional rectangular-duct flow},}\ }\href@noop
  {} {\bibfield  {journal} {\bibinfo  {journal} {Journal of Fluid Mechanics}\
  }\textbf {\bibinfo {volume} {782}},\ \bibinfo {pages} {368--379} (\bibinfo
  {year} {2015})}\BibitemShut {NoStop}%
\bibitem [{\citenamefont {Lemoult}\ \emph {et~al.}(2014)\citenamefont
  {Lemoult}, \citenamefont {Gumowski}, \citenamefont {Aider},\ and\
  \citenamefont {Wesfreid}}]{LemoultEPJE2014}%
  \BibitemOpen
  \bibfield  {author} {\bibinfo {author} {\bibfnamefont {Gr{\'e}goire}\
  \bibnamefont {Lemoult}}, \bibinfo {author} {\bibfnamefont {Konrad}\
  \bibnamefont {Gumowski}}, \bibinfo {author} {\bibfnamefont {Jean-Luc}\
  \bibnamefont {Aider}}, \ and\ \bibinfo {author} {\bibfnamefont
  {Jos{\'e}~Eduardo}\ \bibnamefont {Wesfreid}},\ }\bibfield  {title} {\enquote
  {\bibinfo {title} {Turbulent spots in channel flow: An experimental study},}\
  }\href@noop {} {\bibfield  {journal} {\bibinfo  {journal} {The European
  Physical Journal E}\ }\textbf {\bibinfo {volume} {37}},\ \bibinfo {pages}
  {1--11} (\bibinfo {year} {2014})}\BibitemShut {NoStop}%
\bibitem [{\citenamefont {Borrero-Echeverry}\ \emph {et~al.}(2010)\citenamefont
  {Borrero-Echeverry}, \citenamefont {Schatz},\ and\ \citenamefont
  {Tagg}}]{BorreroPRE2010}%
  \BibitemOpen
  \bibfield  {author} {\bibinfo {author} {\bibfnamefont {Daniel}\ \bibnamefont
  {Borrero-Echeverry}}, \bibinfo {author} {\bibfnamefont {Michael~F}\
  \bibnamefont {Schatz}}, \ and\ \bibinfo {author} {\bibfnamefont {Randall}\
  \bibnamefont {Tagg}},\ }\bibfield  {title} {\enquote {\bibinfo {title}
  {Transient turbulence in {T}aylor-{C}ouette flow},}\ }\href@noop {}
  {\bibfield  {journal} {\bibinfo  {journal} {Physical Review E}\ }\textbf
  {\bibinfo {volume} {81}},\ \bibinfo {pages} {025301} (\bibinfo {year}
  {2010})}\BibitemShut {NoStop}%
\bibitem [{\citenamefont {Avila}\ and\ \citenamefont
  {Hof}(2013)}]{AvilaHof2013}%
  \BibitemOpen
  \bibfield  {author} {\bibinfo {author} {\bibfnamefont {Kerstin}\ \bibnamefont
  {Avila}}\ and\ \bibinfo {author} {\bibfnamefont {Bj{\"o}rn}\ \bibnamefont
  {Hof}},\ }\bibfield  {title} {\enquote {\bibinfo {title} {High-precision
  {T}aylor-{C}ouette experiment to study subcritical transitions and the role
  of boundary conditions and size effects},}\ }\href@noop {} {\bibfield
  {journal} {\bibinfo  {journal} {Review of Scientific Instruments}\ }\textbf
  {\bibinfo {volume} {84}},\ \bibinfo {pages} {065106} (\bibinfo {year}
  {2013})}\BibitemShut {NoStop}%
\bibitem [{\citenamefont {Andereck}\ \emph {et~al.}(1986)\citenamefont
  {Andereck}, \citenamefont {Liu},\ and\ \citenamefont
  {Swinney}}]{Andereck1986}%
  \BibitemOpen
  \bibfield  {author} {\bibinfo {author} {\bibfnamefont {C.~David}\
  \bibnamefont {Andereck}}, \bibinfo {author} {\bibfnamefont {S.~S.}\
  \bibnamefont {Liu}}, \ and\ \bibinfo {author} {\bibfnamefont {Harry~L.}\
  \bibnamefont {Swinney}},\ }\bibfield  {title} {\enquote {\bibinfo {title}
  {Flow regimes in a circular {C}ouette system with independently rotating
  cylinders},}\ }\href@noop {} {\bibfield  {journal} {\bibinfo  {journal}
  {Journal of Fluid Mechanics}\ }\textbf {\bibinfo {volume} {164}},\ \bibinfo
  {pages} {155--183} (\bibinfo {year} {1986})}\BibitemShut {NoStop}%
\bibitem [{\citenamefont {Deguchi}\ and\ \citenamefont
  {Nagata}(2011)}]{DeguchiNagata2011}%
  \BibitemOpen
  \bibfield  {author} {\bibinfo {author} {\bibfnamefont {K.}~\bibnamefont
  {Deguchi}}\ and\ \bibinfo {author} {\bibfnamefont {M.}~\bibnamefont
  {Nagata}},\ }\bibfield  {title} {\enquote {\bibinfo {title} {Bifurcations and
  instabilities in sliding {C}ouette flow},}\ }\href@noop {} {\bibfield
  {journal} {\bibinfo  {journal} {Journal of Fluid Mechanics}\ }\textbf
  {\bibinfo {volume} {678}},\ \bibinfo {pages} {156--178} (\bibinfo {year}
  {2011})}\BibitemShut {NoStop}%
\bibitem [{\citenamefont {Kunii}\ \emph {et~al.}(2016)\citenamefont {Kunii},
  \citenamefont {Ishida},\ and\ \citenamefont {Tsukahara}}]{KuniiICTAM2016}%
  \BibitemOpen
  \bibfield  {author} {\bibinfo {author} {\bibfnamefont {Kohei}\ \bibnamefont
  {Kunii}}, \bibinfo {author} {\bibfnamefont {Takahiro}\ \bibnamefont
  {Ishida}}, \ and\ \bibinfo {author} {\bibfnamefont {Takahiro}\ \bibnamefont
  {Tsukahara}},\ }\bibfield  {title} {\enquote {\bibinfo {title} {Helical
  turbuelnce and puff in transitional sliding {C}ouette flow},}\ }in\
  \href@noop {} {\emph {\bibinfo {booktitle} {Proc. ICTAM 2016, Montr\'eal,
  Canada}}}\ (\bibinfo  {publisher} {IUTAM},\ \bibinfo {year}
  {2016})\BibitemShut {NoStop}%
\end{thebibliography}%

\end{document}